\documentclass[prd,aps,superscriptaddress,showpacs,preprint,preprintnumbers,amsmath,amssymb]{revtex4}
\usepackage{graphicx}% Include figure files
\usepackage{dcolumn}% Align table columns on decimal point
\usepackage{bm}% bold math

\newcommand{\Psibar}{\overline{\Psi}}
\newcommand{\qbar}{\overline{q}}

\def\mpi2{m_\pi^2}
\def\mK2{m_K^2}

\def\mres{m_{\rm res}}
\newcommand{\Dslash}{\rlap{/}\kern-2.0pt D}
\def\Hw{\gamma_5D_w(-M_5)}
\def\HwDag{D^{\dagger}_w(-M_5)}

\newcommand{\bea}{\begin{eqnarray}}
\newcommand{\eea}{\end{eqnarray}}
\newcommand{\be}{\begin{equation}}
\newcommand{\ee}{\end{equation}}

\newenvironment{outline}{
\noindent \framebox{Begin \ outline \hspace{5.0in} }
\begin{enumerate}}
{\end{enumerate}
\noindent \framebox{End \ outline \hspace{5.0in} }
\vspace{0.25in}}

\def\simge{%  ``greater than about'' symbol
    \mathrel{\rlap{\raise 0.511ex
        \hbox{$>$}}{\lower 0.511ex \hbox{$\sim$}}}}
\def\simle{%  ``less than about'' symbol
    \mathrel{\rlap{\raise 0.511ex
        \hbox{$<$}}{\lower 0.511ex \hbox{$\sim$}}}}

\begin{document}
\bibliographystyle{apsrev}

%%%%%%%%%%%%%%%%%%%%%%%%%%%%%% Counters %%%%%%%%%%%%%%%%%%%%%%%%%%
%%
%%  Counters to control sections printed and whether outline is
%%  printed.  Set value to 1 to print corresponding material
%%

\newcounter{Outline}
\setcounter{Outline}{0}

\newcounter{Intro}
\setcounter{Intro}{1}

\newcounter{Actions}
\setcounter{Actions}{1}

\newcounter{WilsonMres}
\setcounter{WilsonMres}{1}

\newcounter{Mres}
\setcounter{Mres}{1}

\newcounter{Topology}
\setcounter{Topology}{1}

\newcounter{Scaling}
\setcounter{Scaling}{1}

\newcounter{Conclusions}
\setcounter{Conclusions}{1}

\newcounter{Acknowledgments}
\setcounter{Acknowledgments}{1}

\newcounter{Appendix}
\setcounter{Appendix}{0}

\newcounter{Tables}
\setcounter{Tables}{1}

\newcounter{Figures}
\setcounter{Figures}{1}

%%%%%%%%%%%%%%%%%%%%%%%%%%%%%% TITLEPAGE %%%%%%%%%%%%%%%%%%%%%%%%%%

\preprint{CU-TP-1051, BNL-HET-02/2, RBRC-233}

\title{Domain wall fermions with improved gauge actions}

\author{Y.~Aoki}
\author{T.~Blum}
\affiliation{RIKEN-BNL Research Center, Brookhaven National Laboratory,Upton, NY 11973}

\author{N.~Christ}
\author{C.~Cristian}
\affiliation{Physics Department,Columbia University,New York, NY 10027}

\author{C.~Dawson}
\affiliation{RIKEN-BNL Research Center, Brookhaven National Laboratory,Upton, NY 11973}

\author{T.~Izubuchi\footnote{on leave from Institute of Theoretical Physics, 
Kanazawa University, Ishikawa, Japan}}
\affiliation{Physics Department, Brookhaven National Laboratory,Upton, NY 11973}
\author{G.~Liu}
\author{R.~Mawhinney}
\affiliation{Physics Department,Columbia University,New York, NY 10027}

\author{S. Ohta}
\affiliation{Institute for Particle and Nuclear Studies, KEK, Tsukuba, Ibaraki, 305-0801, Japan}

\author{K.~Orginos}
\affiliation{RIKEN-BNL Research Center, Brookhaven National Laboratory,Upton, NY 11973}

\author{A.~Soni}
\affiliation{Physics Department, Brookhaven National Laboratory,Upton, NY 11973}

\author{L.~Wu}
\affiliation{Physics Department,Columbia University,New York, NY 10027}

\date{\today}

\begin{abstract}
We study the chiral properties of quenched domain wall fermions with
several gauge actions. We demonstrate that the residual chiral symmetry
breaking, which is present for a finite number of lattice
sites in the fifth dimension ($L_s$), 
can be substantially suppressed using
improved gauge actions.  In particular the Symanzik action, the
Iwasaki action, and a renormalization group improved gauge action,
called doubly blocked Wilson (DBW2), are studied and compared to
the Wilson action.
All improved gauge actions studied show a reduction in the additive residual
quark mass, $\mres$. Remarkably, in the DBW2 case 
$\mres$ is roughly two orders of magnitude smaller than the
Wilson gauge action at $a^{-1}=2$ GeV and $L_s=16$. Significant reduction in
$\mres$ is also realized at stronger gauge coupling corresponding to
$a^{-1}=1.3$ GeV. As our numerical investigation indicates,
this reduction is achieved by reducing
the number of topological lattice dislocations present in the gauge field 
configurations.
We also present detailed results for the quenched
light hadron spectrum and the pion decay constant using the DBW2 gauge
action.

\end{abstract}

\maketitle

\newpage

%%%%%%%%%%%%%%%%%%%%%%%%%%%%%% INTRODUCTION %%%%%%%%%%%%%%%%%%%%%%%%%%

\section{Introduction}
\label{sec:intro}

\ifnum\theIntro=1
%%%%%%%%%%%%%%%%%%%%%%%%%%%%  Section %%%%%%%%%%%%%%%%%%%%%%%%%%%%%%%%
%
%  \section{Introduction}
%  \label{sec:intro}
%  File:  intro.tex
%
%%%%%%%%%%%%%%%%%%%%%%%%%%%%%%%%%%%%%%%%%%%%%%%%%%%%%%%%%%%%%%%%%%%%%%

\ifnum\theOutline=1
\begin{outline}
\item Talk about DWF implementation problems
\item Motivate gauge action search (topology changing confs)
\item Refer to RBC and CP-PACS encouraging Iwasaki action experience.
\end{outline}
\fi

Domain wall
fermions~\cite{Kaplan:1992bt,Kaplan:1993sg,Shamir:1993zy,Furman:1995ky}
are expected to provide an implementation of lattice fermions with
exact chiral symmetry, even at a finite lattice spacing.  To achieve
this exact symmetry, an infinite fifth dimension must be introduced and
simulations have been done to explore the limit of large fifth
dimension for both full and quenched
QCD~\cite{Blum:2000kn,AliKhan:2000iv,Aoki:2001su,Chen:2000zu,Izubuchi:2002pq,Izubuchi:2002pt}.
The finite size of the fifth dimension, $L_s$, used in numerical
simulations, produces a small amount of chiral symmetry breaking, which
should go to zero in the limit $L_s\to\infty$.  In practical
implementations the aim is to achieve the smallest chiral symmetry
breaking possible at a given $L_s$, thus minimizing the cost of the
simulation.  Further information about domain wall fermions and their
applications is given in the recent reviews
\cite{Vranas:2000tz,Hernandez:2001yd}.

There have now been several suggestions on how to minimize the
computational cost of domain wall fermions. An obvious way to achieve
this is to make the five-dimensional eigenvectors of the domain wall
fermion operator, which for small eigenvalues should be localized on
the four-dimensional boundaries of the fifth dimension, decay faster in
the fifth dimension. This reduces the mixing between the opposite
chirality modes, which are bound to opposite ends of the fifth
dimension.  Shamir~\cite{Shamir:2000cf} has calculated the
fifth-dimensional decay of the eigenfunctions with zero eigenvalues
using perturbation theory, suggesting a modification of the
four-dimensional component of the domain wall fermion operator to
increase the decay.  This interesting perturbative result may explain
some of the features seen in non-perturbative simulations.  (Of course,
modifications to the domain wall fermion operator may increase the
computational cost by more than the reduction in $L_s$ reduces it.)
Another method of improving domain wall fermions is proposed
in~\cite{Edwards:2000rk,Hernandez:2000iw}.  The basic idea behind these
proposals is to project out the zero modes of the four-dimensional
Hamiltonian describing the propagation in the fifth dimension. As a
result, the localization on the boundaries of the fermionic light modes
is enhanced.

In this paper we systematically examine a different option: the
modification of the gauge action to suppress the finite $L_s$ explicit
chiral symmetry breaking~\cite{Orginos:2001xa}.  
Note that in principle this is a different
criteria from improving the gauge action to achieve better scaling, in
lattice spacing, of physical observables.  We will investigate the
scaling of observables as well, to check that while reducing the
explicit chiral symmetry breaking, we do not distort the approach to
the continuum limit.  It is worth noting that methods which improve
the domain wall fermion operator, like those suggested by Shamir, and
the one investigated here are likely independent of each other, so a
combination of both techniques may lead to even greater efficiency in
domain wall fermion simulations. However, as we will see, our approach
obviates the need of separately treating the near unit eigenvectors of
the transfer matrix, as gauge configurations for which these occur are
suppressed. This has also been studied in~\cite{Hernandez:2002wi}.

The observation that the gauge action can affect significantly the
chiral symmetry of domain wall fermions is not new.  Both the
RBC~\cite{Wu:1999cd} and CP-PACS~\cite{AliKhan:1999zn,AliKhan:2000iv}
collaborations have observed that the use of the Iwasaki
action~\cite{Iwasaki:1983ii} substantially improves chiral symmetry in
quenched simulations. Also in~\cite{Jung:2000fh} it was observed that
the 1-loop Symanzik~\cite{Alford:1995hw} improved gauge action improves
chiral symmetry to a lesser degree.  Here we extend these results and
explore the reason behind the observed improvement.

This paper is organized as follows.  In section~\ref{sec:actions} we
give a brief description of the gauge actions under study.  In
section~\ref{sec:wilsonmres} we introduce the observables used for
studying chiral symmetry breaking and also present the standard Wilson
action results to provide a reference point.  Section~\ref{sec:mres}
contains results for the different actions we studied. We find that the
doubly blocked Wilson (DBW2) action
\cite{Takaishi:1996xj,deForcrand:1999bi} gives residual chiral symmetry
breaking two orders of magnitude smaller than the Wilson gauge action
at comparable lattice spacings and values of $L_s$.  In
section~\ref{sec:topology} we discuss the dominant mechanism of
explicit chiral symmetry breaking in domain wall fermions, which we
find is driven by lattice artifacts, or dislocations, at the lattice
spacings considered.  These dislocations occur as the topological
charge of the gauge field configuration changes during Monte Carlo
evolution.  Given this large improvement in residual chiral symmetry
breaking and the fact that the DBW2 action has not been used before
with domain wall fermions, in Section~\ref{sec:scaling} we present
results for some hadronic observables in order to confirm consistency
with quenched simulations using other gauge actions, to check scaling
with lattice spacing and to lay a foundation for future 
work~\cite{Aoki:2001dx}.

\fi

%%%%%%%%%%%%%%%%%%%%%%%%%%%%  Section %%%%%%%%%%%%%%%%%%%%%%%%%%%%%%%%

\section{Pure Gauge Lattice Actions}
\label{sec:actions}

\ifnum\theActions=1
%%%%%%%%%%%%%%%%%%%%%%%%%%%%  Section %%%%%%%%%%%%%%%%%%%%%%%%%%%%%%%%
%
%  \section{Pure Gauge Lattice Actions}
%  \label{sec:actions}
%  File:  actions.tex
%
%%%%%%%%%%%%%%%%%%%%%%%%%%%%%%%%%%%%%%%%%%%%%%%%%%%%%%%%%%%%%%%%%%%%%%

\ifnum\theOutline=1
\begin{outline}
\item List actions definitions and original motivations
\item The actions are Wilson, Symanzik
\item Iwasaki : cite RBC dynamical results / problems
\item DBW2: Talk about the phase diagram
\end{outline}
\fi

%%%%%%%%%%%%%%%%%%%%%%%%%%%%  Subsection %%%%%%%%%%%%%%%%%%%%%%%%%%%%%%

As mentioned, we study the chiral properties of quenched domain wall
fermions with Symanzik, Iwasaki, and DBW2 gauge actions.  These actions
are built from closed loops of up to six links and provide a
sample of typical lattice actions used to improve scaling of
observables.  As a baseline for comparisons we start with the Wilson
action~\cite{Wilson:1974sk} which is defined by
\begin{equation}
S_G[U] = - \frac{\beta}{3} \sum_{x;\mu<\nu} P[U]_{x,\mu\nu}
\end{equation}
where $P[U]_{x,\mu\nu}$ is the real part of the trace of the path
ordered product of links around the $1\times 1$ plaquette in the
$\mu,\nu$ plane at point $x$ and $\beta \sim 1/g^2_0$ with $g_0$ the
bare gauge coupling.  This is the original non-abelian gauge action
introduced by Wilson, which has ${\cal O}(a^2)$ errors ($a$ is the
lattice spacing).

To begin, we  study the Symanzik one loop improved
action~\cite{Alford:1995hw} where both ${\cal O}(a^2)$ and ${\cal
O}(g^2a^2)$ errors are removed. This action is defined as
\begin{equation}
S_G[U] = - \frac{\beta}{3} \left(  c_0 \sum_{x;\mu<\nu} P[U]_{x,\mu\nu}
        + c_1 \sum_{x;\mu\neq\nu} R[U]_{x,\mu\nu} 
        + c_2 \sum_{x;\mu<\nu<\sigma}  C[U]_{x,\mu\nu\sigma}
  \right)
\end{equation}
where $R[U]_{x,\mu\nu}$ and $C[U]_{x,\mu\nu\sigma}$ denote the real
part of the trace of the ordered product of SU(3) link matrices along
$1\times 2$ rectangles in the $\mu,\nu$ plane and the
$\mu,\nu,\sigma,-\mu,-\nu,-\sigma$ paths, respectively.  The
coefficients $c_0$, $c_1$, and $c_2$ are computed in tadpole improved
one loop perturbation theory~\cite{Alford:1995hw}.  For this action and
the remaining ones, $\beta \sim 1/g^2_0$ as for the Wilson action, but
the precise numerical factors differ.

In addition to the above actions we also studied the
Iwasaki~\cite{Iwasaki:1983ii} action and the DBW2
action~\cite{Takaishi:1996xj,deForcrand:1999bi}. These actions are both
renormalization group (RG) improved actions in a truncated,
two-parameter space.  They can be written down as
\begin{equation}
S_G[U] = - \frac{\beta}{3}\left(  (1-8\,c_1) \sum_{x;\mu<\nu} P[U]_{x,\mu\nu}
        + c_1 \sum_{x;\mu\neq\nu} R[U]_{x,\mu\nu}\right)
\label{eq:rgaction}
\end{equation}
with $c_1 = -0.331$ for the Iwasaki action and $c_1 = -1.4069$ for the
DBW2 action. In the case of the Iwasaki action the coefficient $c_1$
is computed in  weak coupling perturbation theory.  For the DBW2
action $c_1$ is computed~\cite{Takaishi:1996xj} non-perturbatively
using Swendsen's blocking and the Schwinger-Dyson method.
QCD-TARO has studied~\cite{deForcrand:1999bi} the RG flow in the two
parameter space of the plaquette and the rectangle couplings and
concluded that DBW2 is a good approximation to the RG flow in this
plane at least for a range of coarse lattice spacings.

Although the Iwasaki and DBW2 actions are motivated by the desire to
remain on the RG trajectory for quenched QCD, the truncation to the
explicit form used is an approximation.  The accuracy with which these
truncated actions preserve the RG trajectory must be investigated
numerically.  Simulations with the  Iwasaki action
\cite{Iwasaki:1997sn} and the DBW2 action \cite{deForcrand:1999bi} show
improved scaling of the heavy quark potential and the critical
temperature for the finite temperature phase transition, compared to
the Wilson gauge action.  These actions serve as useful starting
points for studying the effects of the gauge action on residual
chiral symmetry breaking in domain wall fermions.

\fi

%%%%%%%%%%%%%%%%%%%%%%%%%%%%  Section %%%%%%%%%%%%%%%%%%%%%%%%%%%%%%%%

\section{Explicit Chiral Symmetry Breaking with Domain Wall Fermions}
\label{sec:wilsonmres}

\ifnum\theWilsonMres=1
%%%%%%%%%%%%%%%%%%%%%%%%%%%%  Section %%%%%%%%%%%%%%%%%%%%%%%%%%%%%%%%
%
%  \section{Explicit Chiral Symmetry Breaking with Domain Wall Fermions}
%  \label{sec:wilsonmres}
%  File:  wilson_mres.tex
%
%%%%%%%%%%%%%%%%%%%%%%%%%%%%%%%%%%%%%%%%%%%%%%%%%%%%%%%%%%%%%%%%%%%%%%

\ifnum\theOutline=1
\begin{outline}
\item File:  text\_sections/wilson\_mres.tex
\item Definition of $\mres$ and formalism 
\item Figure of $\mres$ vs. conf. \#
\item Figure with spikes  
\item Check what other data we have for the Wilson action
\item Give some motivation (transfer matrix)
\end{outline}
\fi

%%%%%%%%%%%%%%%%%%%%%%%%%%%%  Subsection %%%%%%%%%%%%%%%%%%%%%%%%%%%%%%

The central idea behind domain wall fermions is that four-dimensional
fermionic states of opposite chirality are localized dynamically on
opposite boundaries of an extra fifth dimension. The domain wall
fermions are coupled to four-dimensional gauge fields replicated in the
fifth direction, so the light states can be used to simulate a vector
gauge theory like QCD. The five-dimensional fermion action is a
generalization of the Wilson fermion action~\cite{Wilson:1974sk}
with open boundary conditions in the fifth
dimension~\cite{Shamir:1993zy}. In the free field limit, localization
of a single fermionic flavor on the four-dimensional boundaries occurs
if the five-dimensional fermion mass $M_5$ is in the interval (0,2).
This interval is shifted when interactions are switched on.  For an
infinite fifth dimension ($L_s\rightarrow \infty$), chiral symmetry of
the light states is manifest since they have no overlap.
Four-dimensional light quark $q,\qbar$ fields are constructed from the
five-dimensional fermions $\Psi,\overline{\Psi}$ by
\begin{eqnarray}
 q(x) & = & P_L \Psi(x,0) + P_R \Psi(x,L_s-1)\\
 \qbar(x) & = & \overline{\Psi}(x,L_s-1) P_L + \overline\Psi(x,0) P_R,
\label{eq:qfield}
\end{eqnarray}
where $P_{R/L} = \frac{1}{2} (1 \pm \gamma_5)$ are the right-handed and
left-handed projection operators.
Hence a four-dimensional mass term $m_f\qbar q$ can be introduced
if  the fifth dimension boundaries
are coupled directly with a parameter $m_f$\cite{Shamir:1993zy}.
For finite $L_s$ explicit chiral symmetry breaking 
is induced by the mixing of the light states which now extend across the 
fifth dimension. Our conventions throughout this paper
are the same as those in~\cite{Blum:2000kn}.

In order to quantify the explicit chiral symmetry breaking
induced at finite $L_s$, we define the residual mass ($m_{res}$)
through  the Ward-Takahashi identity~\cite{Furman:1995ky}:
\begin{equation}
  \Delta_\mu \langle {\cal A}^a_\mu(x) O(y) \rangle = 
	2m_f \langle J^a_5(x) O(y) \rangle + 2 \langle J^a_{5q}(x) O(y)
	\rangle + i \langle \delta^a O(y) \rangle,
\label{eq:ward_tak_id}
\end{equation}
where
\begin{equation} 
  {\cal A}^a_\mu(x) = \sum_{s = 0}^{L_s - 1 } {\rm sign}
    \left( s  - \frac{L_s - 1} {2} \right) j^a_\mu(x,s)
  \label{eq:axial_cc}
\end{equation}
is a four-dimensional partially-conserved 
axial current which is constructed from the
five-dimensional conserved vector current,
\begin{equation}
  j^a_\mu(x, s) = \frac{1}{2} \left[ \overline{\Psi}(x+\hat{\mu}, s )
    (1 + \gamma_\mu ) U^\dagger_{x+\hat{\mu}, \mu} t^a \Psi(x,s)
    - \overline{\Psi}(x,s)(1 - \gamma_\mu) U_{x,\mu} t^a
    \Psi(x+\hat{\mu}, s) \right].
\end{equation}
The flavor matrices are normalized to obey 
${\rm Tr}\,(t^a t^b ) = \delta^{ab}$,
$\Delta_\mu f(x) = f(x) - f(x-\hat{\mu}) $ is a simple finite
difference operator, and the pseudoscalar density $J^a_{5}(x)$ is
\begin{eqnarray}
  J^a_{5}(x) & = &
	       - \Psibar(x, L_s - 1)  P_{\rm L} t^a \Psi(x, 0)
	       + \Psibar(x, 0)  P_{\rm R} t^a \Psi(x, L_s-1).
             %    \nonumber \\
	     %& = & \overline{q}(x) t^a \gamma_5 q(x).
                 \label{eq:ward_tak_mass_term}
\end{eqnarray}
Note that $J^a_{5}(x)$ is a four-dimensional pseudoscalar density
constructed from fields on the boundaries of the fifth dimension.
The identity (\ref{eq:ward_tak_id}) differs from the
continuum expression by the presence of the $J^a_{5q}(x)$ term.
$J^a_{5q}(x)$ is analogous to $J^a_{5}(x)$, but is built from fields in
the bulk at
$L_s/2$ and $L_s/2 - 1$.
\begin{equation}
  J^a_{5q}(x) = - \Psibar(x, L_s/2-1)  P_{\rm L} t^a \Psi(x, L_s/2)
		+ \Psibar(x, L_s/2)  P_{\rm R} t^a \Psi(x, L_s/2-1).
  \label{eq:midpt_term}
\end{equation}

We refer to this term as the ``mid-point'' contribution to the
divergence of the axial current. In the low energy limit the effect
of this explicit chiral symmetry breaking is described by a simple added
residual mass term $\mres$ 
so that in this limit $J^a_{5q} \sim \mres J^a_5$~\cite{Blum:2000kn}.
Thus, from the mid-point term we 
define the ratio
\begin{equation}
R(t) =\frac{\sum_{x,y}\langle J^a_{5q}(y,t)J^a_5(x,0)\rangle}
           {\sum_{x,y}\langle J^a_5(y,t)J^a_5(x,0)\rangle},
\label{eq:Mres_t}
\end{equation}
which, for $t$ greater than some $t_{\rm min}$ should be independent
of $t$ and equal to the residual mass, giving
\begin{equation}
\mres =\left.\frac{\sum_{x,y}\langle J^a_{5q}(y,t)J^a_5(x,0)\rangle}
 {\sum_{x,y}\langle J^a_5(y,t)J^a_5(x,0)\rangle}\right|_{t \ge t_{min}}.
\label{eq:Mres}
\end{equation}
As we will see, in our numerical simulations $R(t)$ is essentially
$t$ independent for $t \simge 5$ and the $t$ dependence for $t \simle 5$
will be discussed in Section \ref{sec:mres}.
To calculate $\mres$, we average
over a suitable plateau where $R(t)$ is constant.  In the
subsequent discussion $\mres$ serves as our basic measure of chiral
symmetry breaking. In addition, it is useful to define the ratio
\begin{equation}
r_{[U]}(t) =  \frac{ \sum_{x,y} \langle J^a_{5q}(y,t)J^a_5(x,0)\rangle_{[U]}}
             { \sum_{x,y} \langle J^a_5(y,t)J^a_5(x,0)\rangle_{[U]}},
\label{eq:cnfMres}
\end{equation}
and
\begin{equation}
\bar{r}_{[U]} = \sum_t r_{[U]}(t)
\label{eq:sum_R}
\end{equation}
which are both measures of chiral symmetry breaking on a given gauge
configuration $U$.  

Before presenting results for the improved gauge actions, we discuss
what is known about the Wilson action at $a^{-1} \approx 2$ GeV
($\beta=6.0$). In Fig.~\ref{fig:wilson_mres} we show the residual mass
as a function of $L_s$ (the data are from~\cite{Blum:2000kn}).  While
in perturbation theory $\mres$ is expected to decay exponentially, as
stated in~\cite{Blum:2000kn} the data do not support this. However, its
behavior can be fit with two exponentials with a rather weak decay in
the large $L_s$ limit.  Thus, to decrease $\mres$ by an order of
magnitude we need to increase $L_s$ by a large factor, perhaps of
$O(10)$.

Since $\mres$ is determined by the fifth-dimensional falloff of the
boundary states, decreasing $\mres$ requires improving the falloff.
Analytic arguments have shown that for gauge field satisfying a
smoothness condition, exponential falloff is assured
\cite{Hernandez:1998et,Neuberger:1999pz}.  It is expected that at weak
enough couplings, such a smoothness condition is satisfied, which is
not the case for Wilson gauge lattices at $\beta = 6.0$.  Since the
falloff in the fifth dimension can be related to eigenvalues of an
appropriately defined transfer matrix, $\cal{T}$, in the fifth
dimension, studies \cite{Edwards:1999bm} of the spectrum of the
$\cal{T}$ for Wilson gauge action have been done.  They find a
non-vanishing density of unit or near unit eigenvalues of $\cal{T}$,
showing that undamped propagation in the fifth dimension occurs.
We will also study the spectrum of $\cal{T}$, using gauge configurations
generated with the Wilson, Symanzik, Iwasaki and DBW2 actions.

The transfer matrix $\cal{T}$ \cite{Borici:1999da} is defined by
\begin{equation}
{\cal T} = \frac{1-{\cal H}_t}{1+{\cal H}_t} 
\label{eq:trans_mat}
\end{equation}
with
\begin{equation}
 {\cal H}_t = \frac{1}{2+\HwDag}\Hw
\label{eq:trans_hamil}
\end{equation}
being the Hamiltonian for propagation in the fifth dimension and
$D_w(m)$ being the four-dimensional Wilson Dirac operator.  Following
\cite{Edwards:1999bm} we calculate the eigenvalue spectrum of the
Hermitian Wilson Dirac operator $\Hw$ as a function of $M_5$ (the
so-called spectral flow). From Eq.~\ref{eq:trans_hamil} one sees that a
zero eigenvalue in $\Hw$ corresponds directly to a unit eigenvalue of
the transfer matrix, {\em i.e.} the existence of a five-dimensional
mode that is not damped in the fifth dimension.  In addition,
the number of zeros in the spectral flow determines the index of the
domain wall fermion operator and hence serves as a definition of
topology on the lattice.  Thus, if one is working at a fixed value for
$M_5$ and a gauge field is generated via Monte Carlo which has a unit
eigenvalue of $\cal{T}$, an undamped mode in the fifth dimension occurs
on that configuration.  This configuration is one where we informally
say that topology is changing (in the Monte Carlo update).

When studying the spectral flow on a given configuration, if the flow
approaches the $M_5$-axis, we expect the left and right domain wall
modes to become delocalized leading to mixing and attendant chiral
symmetry breaking.  On the other hand, if there is a large vertical gap
in the spectral flow for values of $M_5$ we use in our simulations, the
chiral modes should remain localized on the boundaries.  In
Fig.~\ref{fig:wilson_flow} we present the spectral flow of the lowest
fifteen eigenvalues for some representative Wilson gauge action
configurations.  Many crossings of the $M_5$ axis are evident and even
the modes that do not cross are not far away from the axis, compared with
the large gap that appears for $M_5 < 0.8$.  
Note that $M_5\sim 0.8$
corresponds to the usual critical mass for Wilson fermions where chiral
symmetry is restored at this gauge coupling($\beta=6.0$).  As we will
see, this picture leads to a relatively large value of $\mres$ for the
Wilson gauge action, though we emphasize that the chiral symmetry
breaking is still very small compared to standard Wilson fermions at
this gauge coupling. In Fig.~\ref{fig:four_wilson_spikes} the ratio
$r_{[U]}(t)$ defined in Eq.~\ref{eq:cnfMres} is plotted  for the same
configurations as in Fig.~\ref{fig:wilson_flow}. The panels in
Fig.~\ref{fig:wilson_flow} and Fig.~\ref{fig:four_wilson_spikes} are in
one to one correspondence.  In the figures, $r_{[U]}(t)$ is quite
dependent on $t$, with large fluctuations occuring over a small range
of $t$.  Since we can see multiple crossings in the
spectral flow, which implies undamped modes in the fifth dimension,
and multiple spikes of $r_{[U]}(t)$ it is natural to investigate
whether these are different manifestations of the same phenomena.

In Fig.~\ref{fig:spikes}(a) we present $\bar{r}_{[U]}$ as a function of
configuration number. It is clear that $\bar{r}_{[U]}$ fluctuates
widely, indicating that there are configurations with larger chiral
symmetry breaking and others with relatively small breaking. The number
of configurations with enhanced chiral symmetry breaking is significant
($\sim 50\%$), consistent with the known result that the transfer
matrix has an appreciable number of near unit
eigenvalues~\cite{Edwards:1999bm}. In addition,
Fig.~\ref{fig:four_wilson_spikes} suggests a close correlation between
configurations showing these spikes and those with crossings in the
spectral flow near $M_5=1.8$.

In order to further examine the nature of chiral symmetry breaking on a
given configuration we take a closer look at the ratio $r_{[U]}(t)$
defined in Eq.~\ref{eq:cnfMres}. In Fig.~\ref{fig:four_wilson_spikes}
and Fig.~\ref{fig:one_spike}(a) we present this ratio for typical
Wilson gauge action configurations, again at $a^{-1} \approx 2$ GeV.
As we can see the dominant part of chiral symmetry breaking comes from
localized regions in time. In particular for the configuration of
Fig.~\ref{fig:one_spike}(a), the Hermitian Wilson Dirac operator has
two small eigenvalues whose eigenvectors are localized around the peaks
of $r_{[U]}(t)$. In addition, very localized peaks of a topological
charge density, constructed as described
in~\cite{DeGrand:1998de,DeGrand:1998ss}, are found around the peaks of
$r_{[U]}(t)$. For further discussion see~\cite{Izubuchi:2002pq}.  Thus
we see that the large fluctuations in $r_{[U]}(t)$ as a function of
$t$ are due to localized gauge field fluctuations which are changing
the topology of the lattice.  This is a crucial observation in
understanding why improved gauge actions can reduce chiral symmetry
breaking for domain wall fermions.

If an improved action can reduce lattice artifact configurations which
are undergoing topology change, then $\mres$ can be reduced.  The
effect of the gauge action on dislocations can be understood by
examining its effects on the classical minima of the action {\it i.e.}
instantons.  Using the results of~\cite{GarciaPerez:1994ki} we can see
that for the Iwasaki and the DBW2 action the
${\mathcal{O}}(a^2/\rho^2)$ correction to the action of an isolated
lattice instanton is positive, hence instantons of small size $\rho$
are suppressed. On the contrary for the Wilson action the
${\mathcal{O}}(a^2/\rho^2)$ correction is negative, consequently the
small lattice instantons are enhanced.  This suggests that for the
Iwasaki and the DBW2 actions, gauge configurations with very localized
concentrations of topological charge density are suppressed.  If in
addition, there is a suppression of configurations where localized
topology change is occuring, there will be a reduction of explicit
chiral symmetry breaking.  In conclusion, configurations of non-zero
topology do not produce large residual chiral symmetry breaking, only
configurations where topology is changing, {\em i.e.} where the
spectral flow has a zero.  Suppression of lattice artifact toplogy
changing configurations should decrease $\mres$.

\fi

%%%%%%%%%%%%%%%%%%%%%%%%%%%%  Section %%%%%%%%%%%%%%%%%%%%%%%%%%%%%%%%

\section{Chiral Symmetry with Improved Gauge Actions}
\label{sec:mres}

\ifnum\theMres=1
%%%%%%%%%%%%%%%%%%%%%%%%%%%%  Section %%%%%%%%%%%%%%%%%%%%%%%%%%%%%%%%
%
%  \section{Chiral Symmetry with Improved Gauge Actions}
%  \label{sec:mres}
%  File:  mres.tex
%
%%%%%%%%%%%%%%%%%%%%%%%%%%%%%%%%%%%%%%%%%%%%%%%%%%%%%%%%%%%%%%%%%%%%%%

\ifnum\theOutline=1
\begin{outline}
\item My lat01 proceedings
\item Show the $\mres$ results for all the actions tested
\item What do I do about the hadron mass dependence on $L_s$?
      I think it should be added in the scaling section.
\end{outline}
\fi

%%%%%%%%%%%%%%%%%%%%%%%%%%%%  Subsection %%%%%%%%%%%%%%%%%%%%%%%%%%%%%%

In order to study the effects of the choice of the gauge action on 
 the residual chiral symmetry breaking we performed a series of quenched
simulations using the Symanzik, Iwasaki and DBW2
actions.  In all cases the lattices were $16^3\times 32$ with
inverse lattice spacing $a^{-1} \approx 2$ GeV. We used the $\rho$ mass
to set the scale but also confirmed consistency with the
scale set from the string tension; both yield
equal lattice spacings to within a few percent.  The
mass $M_5$  was tuned to be optimum with an accuracy of
about 5\%. 
%From the point of view of the transfer matrix, 
%the optimum value of $M_5$ is the one that
%has the largest gap (neglecting crossings) in the spectral flow of the
%Hermitian Wilson Dirac operator, and hence the smallest value of
%$\mres$ at fixed $L_s$.
Simulations on a few configurations at several values of
$M_5$ were all that were needed for this determination.
It turns out that for all actions at $a^{-1}\approx 2$ GeV 
the optimum value is roughly 1.8,
except for the DBW2 for which it is 1.7. In the free field limit, the
optimum value is  $M_5=1$\cite{Shamir:1993zy}.
The bare quark masses in our study ranged
from $m_f=0.010$ to $0.060$. A summary of the simulation parameters
is presented in Table~\ref{tab:mres_sim_param}.

 The residual mass was extracted by fitting to a constant at
large time separations the ratio defined in Eq.~\ref{eq:Mres}. Errors are determined by the jackknife method.  As it can be 
seen in Fig.~\ref{fig:mres_vs_t}, this ratio exhibits a fairly stable
plateau at time separations larger than five or six, so we chose a
fitting range of 7-16 in all cases. All data in this figure are for
$L_s=16$ and for bare quark mass 0.020. The quark mass dependence of
the residual mass is mild as seen in Fig.~\ref{fig:mres_vs_mq}.  Since
we have also matched the lattice spacings, it is safe to compare all
the actions at the same bare quark mass ignoring renormalization effects.
Because the numbers we are comparing differ by orders of magnitude these
effects can be safely neglected. In fact the multiplicative
 quark mass renormalization
constants have been computed~\cite{Dawson:2002nr}
and shown to be equal within 5\%.
 In order to eliminate some of
the effects of the remaining small mismatch of the lattice spacings,
we have plotted the residual mass  scaled by the square root of the string
tension.

In Fig.~\ref{fig:mres_vs_ls} we present our measurements of $\mres$
for each action for several values of
$L_s$. In the case of the Iwasaki action we only performed the
measurement at $L_s=16$, and our result agrees with that of
CP-PACS~\cite{AliKhan:2000iv}. The remaining Iwasaki
points are from the CP-PACS publication\cite{AliKhan:2000iv}. 
As one can see at $L_s=16$,
the DBW2 residual mass is about two orders of magnitude smaller than
the residual mass of the Wilson action while the Iwasaki residual mass is
about an order of magnitude smaller than that of the Wilson action.
Finally, the residual mass of the Symanzik action is
roughly a factor of three smaller than that of the Wilson action.  In
this figure the solid lines represent fits to simple exponentials in
all cases except the Wilson action where a fit to two exponentials
is shown. For the Symanzik data a
small deviation from the simple exponential fit is observed at
$L_s=16$ while the Wilson action shows a very clear deviation.
On the contrary, both the Iwasaki and DBW2 data can
be fit well with a simple exponential for the same range of
$L_s$.  
For that reason it is interesting to quote a value for the
parameter $q$ that Shamir has computed
perturbatively~\cite{Shamir:2000cf}. 
His one loop result is that the light fermion wave function $\chi(s)$
decays exponentially away from the wall, i.e $\chi(s) \sim q^s$ with
$q=\frac{1}{2}$. The residual mass also behaves as $\mres \sim
q^{L_s}$.  
In the  case of the Wilson and possibly the Symanzik action, 
the fact
that no good fit to a single exponential is obtained may be a signal that
$m_{res}$ scales as a power law~\footnote{We thank Y. Shamir for
discussions on this point.}, and 
$q\sim 1$. Such behavior is
consistent with the spectral flows observed for the Wilson gauge action.
For the Iwasaki and DBW2 actions $q\approx 0.7$
and $q\approx 0.6$, respectively, which is consistent with a 
gap in the spectral flow at $M_5=1.7-1.8$ that is
well defined on most configurations. We come back to this point in the
following where we investigate the spectral flow for each gauge action.

Given the dramatic improvement in $\mres$ for the DBW2 action, it is
natural to wonder whether further improvement is possible.  We have
explored simulations where the coefficients of the plaquette and
rectangle term in Eq. \ref{eq:rgaction} take on various ratios and
found that for ratios not far from the DBW2 choice, small lattice
spacings could not be achieved.  In addition, a double peaked
distribution of the plaquette values could be found by changing
the ratio of the plaquette and rectangle coeffecients to be about
a factor of 2 different than for the DBW2 action.  Thus, further
dramatic improvement in $\mres$ does not seem possible with an action
which involves only plaquette plus rectangle terms.

\fi

%%%%%%%%%%%%%%%%%%%%%%%%%%%%  Section %%%%%%%%%%%%%%%%%%%%%%%%%%%%%%%%

\section{Topology and chiral symmetry breaking}
\label{sec:topology}

\ifnum\theTopology=1
%%%%%%%%%%%%%%%%%%%%%%%%%%%%  Section %%%%%%%%%%%%%%%%%%%%%%%%%%%%%%%%
%
%  \section{Topology and Chiral Symmetry Breaking}
%  \label{sec:topology}
%  File:  topology.tex
%
%%%%%%%%%%%%%%%%%%%%%%%%%%%%%%%%%%%%%%%%%%%%%%%%%%%%%%%%%%%%%%%%%%%%%%

\ifnum\theOutline=1
\begin{outline}
\item Spikes of $mres$
\item dirty lego plots
\item Wilson flows?
\item Correlations between all this in order to make clear
      the role for topology changing configurations
\item slow topology change for DBW2
\end{outline}
\fi

%%%%%%%%%%%%%%%%%%%%%%%%%%%%  Subsection %%%%%%%%%%%%%%%%%%%%%%%%%%%%%%

In this section we take a closer look at how the different gauge
actions affect explicit chiral symmetry breaking in domain wall
fermions.  As mentioned before, in Fig.~\ref{fig:spikes}(a) the
quantity $\bar{r}_{[U]}$ defined in Eq.~\ref{eq:sum_R} is presented as
a function of the configuration number. The large fluctuations (spikes)
indicate that there are configurations with relatively large chiral
symmetry breaking and configurations with relatively small breaking.
The configurations with large spikes are those for which the transfer
matrix in the fifth dimension has a near unit eigenvalue, or a
corresponding (near) zero eigenvalue of the hermitian Wilson Dirac
operator. In those cases that we have checked for the Wilson gauge
action, a spike is always accompanied by a localized (near) zero
eigenvector of the the Wilson Dirac operator. In addition, the fact
that the spectral flows presented in Fig.~\ref{fig:wilson_flow} have so
many crossings very close to the simulation point $M_5=1.8$ is
consistent with the large number of spikes in
Fig.~\ref{fig:spikes}(a).  In configurations where a spike does not
occur, i.e. no crossing close to $M_5=1.8$, the chiral symmetry
breaking is controlled by the size of the gap of the bulk modes in the
spectral flow. Here we are separating the small Wilson Dirac
eigenvalues into two groups: those that cross zero near $M_5$ and those
that form a more continuum band which we refer to as bulk modes.  In
the case of the Wilson action and configurations with no crossings
close to $M_5$, the bulk mode gap is rather small and not very well
defined; thus even on these configurations the chiral symmetry breaking
is relatively large for a given $M_5$.

For the Symanzik action (Fig.~\ref{fig:spikes}(b)) the number of spikes
is slightly smaller than in the case of the Wilson action, and also the
number of crossings in the spectral flow (Fig.~\ref{fig:Symanzik_flow})
is correspondingly reduced. Also, the bulk mode gap is larger.  As a
result the baseline, or level of the troughs between peaks in
$\bar{r}_{[U]}$, is lower than in the case of the Wilson action,
contributing to the reduction in the residual mass.

The above picture becomes much clearer with the Iwasaki
(Fig.~\ref{fig:spikes}(c)) and  DBW2 actions
(Fig.~\ref{fig:spikes}(d)). The number of spikes is significantly
smaller, and the baseline is well defined (especially for the DBW2
action).  The typical spectral flows presented in
Fig.~\ref{fig:iwasaki_flow} and Fig.~\ref{fig:dbw2_flow} again support
the fact that the Iwasaki action, and to a larger degree the DBW2
action, significantly suppress the near unity eigenvalues of the domain
wall fermion transfer matrix.  In both cases the gap of the bulk modes
in the spectral flow becomes significantly larger. As a result the
explicit domain wall fermion chiral symmetry breaking is significantly
reduced.

In Fig.~\ref{fig:one_spike} we present the ratio $r_{[U]}(t)$ defined
in Eq.~\ref{eq:cnfMres} for a typical configuration of each action. In
all cases it is evident that the dominant contribution to chiral
symmetry breaking comes from very localized objects, and thus as we
argued before, it is not very surprising that local ${\mathcal O}(a^2)$
modifications of the gauge action can have a very significant effect on
explicit residual chiral symmetry breaking.

It is important to recognize that the above mechanism for explicit
chiral symmetry breaking is related to topology-changing
configurations (see~\cite{Narayanan:1998yv} and references therein for
a more complete discussion).  The connection is made through the index
theorem: the domain wall fermion operator in the limit $L_s
\rightarrow \infty$ has an index~\cite{Narayanan:1994sk,Furman:1995ky}
equal to the number of right- minus the number of left-handed zero
modes, which corresponds to the topological charge of the background
gauge field configuration---a quantity which becomes precise in the
continuum limit.  This integer depends on the value of $M_5$ used and
is given by the net number of crossings in the spectral flow of the
Wilson Dirac operator as the Wilson mass varies between a value above
the critical Wilson mass and $-M_5$.  While this index is well-defined
only in the limit $L_s \rightarrow \infty$, our simulations show that
the near-zero eigenvectors of the finite-$L_s$ operator obey the index
theorem to a high degree of accuracy~\cite{Blum:2000kn,Blum:2001qg}.
In particular, for an Iwasaki $a^{-1} \approx 2$~GeV ensemble, when
compared to the topological charge computed using the smoothing method
described in~\cite{DeGrand:1998ss,DeGrand:1998de}, the index agrees
very well. In those cases where the topological charge is not close to
an integer, we also find a crossing in the spectral flow, a spike in
$r_{[U]}(t)$, and a complex structure of eigenvectors that is not
expected from simple chiral symmetry arguments\cite{Blum:2001qg}.  If
$M_5$ sits exactly on a crossing, then the index is not defined, even
the limit $L_s\to\infty$.  A crossing in the spectral flow that occurs
away from the critical Wilson mass corresponds to a configuration with
indistinct topology.  Put differently, if the particular gauge field
in question is in the midst of changing its topology, which must
happen if the update algorithm is ergodic and updates the
configuration smoothly, then such a gauge field must give rise to a
crossing. It is also sensible that such a tunneling from one
topological sector to another proceeds through local changes in the
gauge field which have a characteristic size of one to two lattice
spacings. In the continuum limit, if the density of these dislocations
is zero, then all crossings happen at the critical mass and correspond
to physical topological charge. Thus the index as computed from the
spectrum of the domain wall operator Dirac operator is well-defined in
this case.

Consequently, when the Iwasaki action or the DBW2 action is used, the
question arises whether the topology changes efficiently.  We have
measured the topological charge using the smoothing method described
in~\cite{DeGrand:1998ss,DeGrand:1998de} \footnote{We thank the MILC
collaboration for their code which was used to compute the topological
charge.}.  Our data are presented in Fig.~\ref{fig:qtop}. The
configurations shown in this figure are separated by 1000 sweeps 
of Cabibo-Marinari pseudo-heatbath with a Kennedy-Pendleton accept/reject 
step.~\footnote{Each sweep consists of one update of two independent
SU(2) subgroups of each SU(3) link.} We can see that there is a
significant slow down in the topological charge fluctuations for the
DBW2 action. Both the Symanzik and the Iwasaki action also show a mild
reduction in the frequency  of change of the topological charge.
Although the problem seems severe for the DBW2 action, we can tackle
it with brute force. For that reason we have produced a library of DBW2
lattices to be used for several domain wall fermion projects.  Given
the considerable cost of measuring domain wall fermionic observables,
this higher cost of producing DBW2 lattices at 2~GeV is negligible.
However, it is clear that this brute force approach will become less
practical for smaller lattice spacing since topology change is likely
to be rapidly suppressed as we approach  the continuum
limit\footnote{Private communication with P. de Forcrand.}.

\fi

%%%%%%%%%%%%%%%%%%%%%%%%%%%%  Section %%%%%%%%%%%%%%%%%%%%%%%%%%%%%%%%

\section{Hadronic observables for the DBW2 action}
\label{sec:scaling}

\ifnum\theScaling=1
%%%%%%%%%%%%%%%%%%%%%%%%%%%%  Section %%%%%%%%%%%%%%%%%%%%%%%%%%%%%%%%
%
%  \section{Hadronic observables and scaling}
%  \label{sec:scaling}
%  File:  scaling.tex
%
%%%%%%%%%%%%%%%%%%%%%%%%%%%%%%%%%%%%%%%%%%%%%%%%%%%%%%%%%%%%%%%%%%%%%%

\ifnum\theOutline=1
\begin{outline}
\item Yasumichi's Lat01
\item $\rho$ and nucleon masses 
\item Decay constants / $Z_A$
\item J-parameter
\item $\bar\psi\psi$ if we get $Z_m$ meanwhile 
\end{outline}
\fi

In this section we discuss various hadronic observables calculated
with the DBW2 gauge action at $\beta=0.87$ and $1.04$ 
which correspond to $a^{-1}\approx 1.3$
and $2$ GeV respectively.

%%%%%%%%%%%%%%%%%%%%%%%%%%%%  Subsection %%%%%%%%%%%%%%%%%%%%%%%%%%%%%%

\subsection{The heavy quark potential}

We measure the heavy quark potential as in~\cite{Bernard:2000gd} by
fixing to Coulomb gauge and then computing the two
point correlation function of products of temporal links. More
precisely,
\begin{equation}
A\;e^{-V(r)T} = \sum_{t_0,\vec{x}} \langle
  {\rm Tr}\left[L_{t_0,T}(\vec{x})L_{t_0,T}^\dagger(\vec{x}+\vec{r})
  \right]\rangle,
\label{eq:potential} 
\end{equation}
with 
\begin{equation}
\label{eq:pot_line}
 L_{t_0,T}(\vec{x}) = \prod_{t=t_0}^{t_0+T} U_t(\vec{x},t),
\end{equation}
and $V(r)$ the heavy quark potential.
The potential is extracted by taking ratios of the correlation
function in Eq.~\ref{eq:potential} at $T$ and $T+1$. 
The systematics involved in choosing $T$ were carefully studied
and the optimal $T$ was chosen. For the 1.3 GeV lattices 
T was 4 while for the 2 GeV lattices it was 7.
The potential $V(r)$ is fit to
\begin{equation}
V(r) = C -\frac{\alpha}{r} + \sigma r.
\label{eq:potential_fit}
\end{equation}

The above formula gave very good fits for spatial distances $r > \sqrt
2$.  The upper range of $r$ was determined by the distance where the
error on the potential became unacceptably large.  The maximum distance
used was $r = 6$ and 7 for the 1.3 and 2 GeV lattices, respectively.
Figs.~\ref{fig:potb087}, \ref{fig:potb104} show the heavy quark
potential as a function of distance. The results for the string tension
$\sigma$ and the Sommer parameter~\cite{Sommer:1994ce,Luscher:1994gh}
$r_0$ are tabulated in Table \ref{tab:DBW2param}. These results are
used in our subsequent discussion of the scaling of hadronic
observables.

\subsection{Simulation and analysis}

For each set of gauge configurations, domain wall fermion propagators
are computed with two types of sources: 1) a local point source and 2)
a Coulomb gauge fixed extended source which is either a wall source for
$\beta=1.04$ or box source with $8^3$ volume for $\beta=0.87$.  (We set
the source to one at each site inside the box and zero elsewhere.)  The
local source is used for the determination of the decay constants and
also the axial current renormalization factor $Z_A$ ($\beta=0.87$
only). The extended source is used for all other purposes.

In Table.~\ref{tab:DBW2mres} we give $\mres$ in the chiral
limit for the same ensemble of configurations used 
for the hadronic observables to be discussed in this section. 
We have fitted $\mres(m_f)$ with a linear function of $m_f$ 
to obtain $\mres(0)$ for which the chiral limit of low energy 
physics is defined as $m_f = -\mres(0)$. All data are used to
extrapolate to $m_f\to 0$ for $\beta=1.04$. On the other hand,
the largest value $m_f=0.09$ for $\beta=0.87$ is not used for the
extrapolation.

We take the chiral limit $m_f\to-\mres(0)$ as the physical point for u, d
quarks. This determines the physical $\rho$ meson mass $m_\rho$.
With the input $m_\rho=770$ MeV, the lattice spacing $a$ is determined.
The kaon physical point $m_f^K$, which roughly corresponds to half the
strange quark mass, for $f_K$ and $m_{K^*}$ is defined by
$m_\pi(m_f=m_f^K)/m_\rho(m_f=-\mres) = 0.645$ using 
only degenerate quark masses. 
We do this procedure for every jackknife 
sample to estimate the error for values at the physical kaon point.

\subsection{Chiral property of pseudoscalar mass}

Because of the almost exact chiral symmetry of domain wall fermions
and the use of the quenched approximation, the pion two-point function
suffers contamination from topological near zero modes, which causes a
shift in fitted masses from their infinite volume
values\cite{Blum:2000kn}.  The effect is expected to be inversely
proportional to the square root of the volume.
Because we used different physical volumes for the two different gauge
couplings ($V\simeq (2.4 {\rm fm})^3$ for $\beta=0.87$,
$(1.6 {\rm fm})^3$ for $\beta=1.04$), 
the size of the effect on the pseudoscalar mass should be
different in our two lattice ensembles.  To study zero mode effects we
examine the pseudoscalar mass from two different two-point functions.
One is the pseudoscalar-pseudoscalar correlator (PP)
\begin{equation}
 \langle \pi^a(x) \pi^a(0) \rangle = 
 \langle iJ_5^a(x)\, i J_5^a(0) \rangle = 
  - \langle\qbar \tau^a \gamma_5q(x)\ \qbar \tau^a \gamma_5q(0)\rangle,
\end{equation}
and the other is the correlator of the temporal component the
axial-vector current (AA)
\begin{equation}
 \langle A_0^a(x) A_0^a(0) \rangle = 
  \langle\qbar \tau^a \gamma_5\gamma_0 q(x)
  \ \qbar \tau^a \gamma_5\gamma_0q(0)\rangle,
\end{equation}
where $q$ and $\qbar$ are four dimensional quark fields defined in 
Eq.~\ref{eq:qfield},
As discussed in~\cite{Blum:2000kn}, the two types of correlators
suffer differently from topological zero modes.
The leading contribution to the pseudoscalar correlator is 
$\sim 1/(m_f^2 \sqrt V)$ while it is $\sim 1/(m_f \sqrt V)$ 
for the axial correlator.  Although the relative contribution 
from the pole compared to the physical one is the same for both, 
the prefactor of the $1/(m_f \sqrt V)$ term is
expected to be suppressed~\cite{Blum:2000kn}. 
Thus, the mass extracted from the PP
correlator is expected to have a stronger finite volume effect from zero modes 
than the AA correlator. The observed effect on the meson mass
calculated for light quark masses is to shift it
above a linear extrapolation from the region of heavier quark mass.

The pseudoscalar mass extracted from both types of correlators
is presented in Table \ref{tab:hadronmass}.
Fig.~\ref{fig:pi-mf_b087} shows the pseudoscalar mass squared as a
function of $m_f$ for $\beta=0.87$. 
Both values of the pion mass are consistent with each other for $m_f\ge0.02$.
However, at $m_f=0.01$ the mass extracted from the PP correlator lies
above the AA one, outside of their statistical errors. 
Because the axial correlator is expected to
have smaller finite volume effects from zero modes, we
use this correlator for further analysis.

The linear fit of the pion mass squared in $m_f$ is quite good 
in the region $0.01\le m_f\le 0.06$ as indicated by the $\chi^2/dof$, which
is tabulated in Table \ref{tab:pi-linearfit}. Note, we are using the
same set of gauge configurations for all values of $m_f$ but employ an
uncorrelated fit. One can reliably
extract the physical kaon mass;
however, the fit overshoots the point
$m_f=-\mres$ where the pion mass should vanish. This is a signal of
non-linearity for the pion mass at small $m_f$. Instead of a linear
function we should employ the quenched chiral log \cite{Bernard:1992mk}  
formula with the constraint that the pion mass vanishes at $m_f=-\mres$, 
\begin{equation}
 m_{\pi}^2  =  a_\pi ( m_f + \mres )
  \left[ 1 - \delta \log \left( \frac{ a_\pi ( m_f + \mres ) }
			  {\Lambda_{Q\chi PT}^2}
			\right) \right] + b ( m_f + \mres )^2,
 \label{eq:chilog}
\end{equation}
where we have introduced the quadratic term in addition to the
expression used in \cite{Blum:2001xb}.
As is discussed in Ref.~\cite{Blum:2001xb}, Eq.~\ref{eq:chilog} has a
form suggested by chiral 
perturbation theory.  Note that the explicit chiral perturbation
theory formula, {\it e.g.} Eq.~90 of Ref.~\cite{Blum:2001xb}, involves three 
parameters $\alpha$, $\delta$ and $\Lambda_{Q\chi PT}$ and additional
terms.   However, within our parameter range and numerical accuracy 
these additional terms have the effect of adding an undetermined, 
$m_f$-independent constant to the expression within the square brackets.  
Our choice of zero for this unknown constant represents a rescaling of 
the parameters $\delta$ and $\Lambda_{Q\chi PT}$ from those that
appear in the chiral perturbation theory prediction.

The data fit this formula well with a reasonable value of
the chiral log coefficient $\delta$ as listed in Table
\ref{tab:chilog}, where the scale is set as $\Lambda_{Q\chi PT}=1$ GeV. 
If the coefficient of the quadratic term $b$ is not set to
zero in the fit, a somewhat larger value of $\delta$ results. This is
because the two terms tend to cancel each other. Note that the value of $b$ 
is the same order as $a_\pi$. Ultimately, a proper covariant fit with
reliable $\chi^2$ should be used to distinguish the two fits. At present
we do not have enough statistics to do this.

Fig.~\ref{fig:pi-mf_b104} shows the pion mass squared for
$\beta=1.04$. 
The results for the larger value of $m_f$ agree with each other, but
deviations appear as $m_f$ decreases.
In fact, pronounced upward curvature of the data, which was not observed
for the axial correlator for $\beta=0.87$,
result in large
values of $\chi^2/dof$ listed in Table \ref{tab:pi-linearfit} for both
correlators. The lines in the Fig.~\ref{fig:pi-mf_b104} show the results
of the fit excluding the lightest point $m_f=0.01$, which makes 
$\chi^2/dof$ smaller. Thus we can assume that the pion mass for $m_f\ge
0.015$ does not suffer much from zero mode effects. As this fit
reproduces the data in the range $0.015\le m_f \le 0.04$ very well,
we use it for the interpolation of the kaon point.
Also using this region of $m_f$, one can fit data with the quenched
chiral log without the quadratic term for $\chi^2/dof=0.5(1)$. The
resulting  $\delta$ is consistent to that for $\beta=0.87$. Since our
range of  $m_f$ for $\beta=1.04$ is not wide enough to disentangle the
quadratic term and the log term, we do not list a result for non-zero $b$.

\subsection{Hadron spectrum}

We list the results for the vector meson and nucleon
masses in Table \ref{tab:hadronmass}. Figs.~\ref{fig:rhoN-mf_b087} 
and \ref{fig:rhoN-mf_b104}
show the vector meson and nucleon masses
as functions of $m_f$ for $\beta=0.87$ and $1.04$ respectively.  
Physical nucleon, rho, and $K^*$ masses are indicated on the figure.

Fig.~\ref{fig:rho-a} presents the rho meson mass versus lattice spacing
squared, both normalized with $r_0$. We also plot the results obtained
with the Wilson gauge action from ref.~\cite{Blum:2000kn}. 
We have selected values 
obtained at $(\beta,V,L_s) = (6,16^3,24), (5.85,12^3,20), (5.7,8^3,48)$.
These lattices have almost the same physical volume, and
$L_s$ is the largest available in each case. 
We observe  consistency
between DBW2 and Wilson actions. The flatness of the data
reflects the small size of the scaling violation.

Fig.~\ref{fig:NKstar-a} is a scaling plot for the nucleon and $K^*$ 
masses normalized by the $\rho$ meson mass. Again good scaling is observed. 
The $K^*$ appears to be lighter than experiment, which is consistent 
with other quenched simulations. For the nucleon mass, there is an 
observed discrepancy between the results with Wilson type fermions 
and staggered fermions (see comparison by S.~Aoki \cite{Aoki:2000kp}). 
The former gives a lighter nucleon mass while the latter is consistent with 
the experiment. The nucleon mass for domain-wall fermion is slightly larger
than experiment for the lattice spacings we examined. We need to perform
simulations for larger physical volume at $a^{-1}\simeq 2$ GeV,
as well as simulations at smaller lattice spacings to do the continuum
extrapolation needed to compare with conventional fermions.
For the comparison within the domain-wall fermions, given the statistics
and the fact that 
the physical volumes of our ensembles are not the same, we can
not say if the DBW2 action exhibits better scaling than the Wilson gauge
action though it seems that the scaling is at least as good.

The J parameter, which is introduced in ref.~\cite{Lacock:1995tq} to
examine the effect of quenching on the mass spectrum, is defined by  
\begin{equation}
% J = m_{K^*}\frac{d m_\rho}{d m_\pi^2}\big|_{m_\rho/m_\pi=m_{K^*}/m_{K}}.
 J = m_{K^*}\frac{d m_\rho}{d m_\pi^2},
\end{equation}
where $m_{K^*}(m_f)$ is evaluated at
$m_\rho(m_f)/m_\pi(m_f)=m_{K^*}/m_K=1.8$ by definition. The slope of
$m_\rho$ with respect to $m_\pi^2$ is determined with the data in
$0.01\le m_f \le 0.06$ for $\beta=0.87$ and $0.015\le m_f \le 0.04$ for
$\beta=1.04$, using pion mass determined with AA.
An approximate phenomenological value calculated from the experimental 
mass spectrum is, 
\begin{equation}
 J = m_{K^*}\frac{m_{K^*}-m_{\rho}}{m_{K}^2-m_{\pi}^2} = 0.48.
\end{equation}
Our lattice results for the J parameter, listed in Table
\ref{tab:rho_fit} and plotted in fig.~\ref{fig:J}, show a significant
difference from the phenomenological value.  However, these results are
consistent with other quenched results
% \cite{Bowler:1999ae,Bernard:2001av}.
(see a summary given by T.~Kaneko \cite{Kaneko:2001ux}).
Note that we have used degenerate quark
masses which could explain some of the difference between our result and 
experiment, but the largest source of the discrepancy is 
likely to be due to quenching.

\subsection{Pseudoscalar decay constants}

The pseudoscalar decay constants are calculated from the amplitude of
the point-point two-point function of the temporal component of local
axial vector current,
\begin{equation}
 \frac{f_\pi^2}{Z_A^2}\frac{m_\pi}{2}e^{- m_\pi t} = 
  \sum_{\vec{x}}\langle A_0^a(\vec{x},t) A_0^a(0)\rangle,
\end{equation}
for $t \gg 1$.
This simple local current is not partially conserved, unlike ${\cal
A_\mu}$ (Eq. \ref{eq:axial_cc}) which obeys the Ward--Takahashi
identity (Eq.~\ref{eq:ward_tak_id}), so the local current receives a
multiplicative renormalization. We can calculate the renormalization
factor $Z_A$ from
the ratio of the correlation functions of the two currents.  Here we
employ a ratio designed to remove some of the lattice spacing
error\cite{Blum:2000kn}.
\begin{equation}
{\cal R}(t) = \frac{1}{2}\left[\frac{C(t+1/2)+C(t-1/2)}{2 L(t)} 
		+ \frac{2 C(t+1/2)}{L(t)+L(t+1)}\right],
\label{eq:R}
\end{equation}
where $C$ and $L$ are the correlators of the pseudoscalar density
with the partially conserved and local axial currents respectively,
\begin{eqnarray}
 C(t+1/2) &=& \sum_{\vec{x}}\langle{\cal A}_0^a(\vec{x},t) P^a(0)\rangle,\\
 L(t) &=& \sum_{\vec{x}}\langle A_0^a(\vec{x},t) P^a(0)\rangle .
\end{eqnarray}
The first and second terms on the r.h.s.\ of Eq.~\ref{eq:R} remove the
${\cal{O}}(a)$ scaling error, and suppress the ${\cal{O}}(a^2)$ error 
in the sum.  The value of $Z_A$ is determined by
fitting ${\cal{R}}(t)$ to a constant for the interval $6\le t\le 26$,
where ${\cal{R}}(t)$ is flat to a very good approximation. The results
are given in Table \ref{tab:fps-mf} and plotted in
Fig.~\ref{fig:Z_A-mf}. An estimate of $Z_A$ with only the first term
is also plotted to demonstrate that the complex ratio in
Eq.~\ref{eq:R} actually works. Although linear dependence on $m_f$ in
$\cal{R}$ remains for both cases, it is very small for the full
expression of Eq.~\ref{eq:R}.  We use the value $Z_A$ obtained in the
chiral limit $m_f\to -\mres$ for the calculation of the pseudoscalar
decay constant, which is given in Table \ref{tab:fps}.

Another method to calculate the decay constant
is to use the Ward--Takahashi identity to derive its relation with the
pseudoscalar to vacuum matrix element of the pseudoscalar density:
\begin{equation}
 \frac{f_\pi}{(m_f+\mres)}\frac{m_\pi^2}{2} = 
  \langle 0| J_5^a|\pi,\vec{p}=\vec{0}\rangle.
\end{equation}
This matrix element is determined from the PP correlator.
Results with each correlator for $\beta=0.87$ are shown in
Fig.~\ref{fig:fps-mf_b087}.
For both methods we use the results of $m_{\pi}$ extracted from the AA
correlator with the extended source quark propagator since it is our
most precise determination.

Consistency between both results shows that we have good control
over the chiral symmetry breaking through our determination of $\mres$.
Also, possible zero mode effects, which influence the two 
correlators differently, appear to be small at least for $m_f\ge0.02$,
which is consistent with the absence of zero mode effects in the pion
mass over the same range of $m_f$.

Fig.~\ref{fig:fps-mf_b104} is the same plot but for $\beta=1.04$.
The results from the two determinations
are also in good agreement.
Results for decay constants at the physical points are given in 
Table~\ref{tab:fps} and plotted in Fig.~\ref{fig:fps-a} against 
lattice spacing squared. 
They are consistent with those reported for the Wilson
action \cite{Blum:2000kn}.
We observe good scaling both for $f_{\pi}$ and $f_{K}$.
The result for $f_{\pi}$ is in good agreement with the experimental value.
However, the ratio $f_K/f_\pi$ appears to be smaller than the
experimental value $1.22$, which is an expected effect of 
quenching\cite{Bernard:1992mk}.

\fi

%%%%%%%%%%%%%%%%%%%%%%%%%%%%  Section %%%%%%%%%%%%%%%%%%%%%%%%%%%%%%%%

\section{Conclusions}
\label{sec:conclusions}

\ifnum\theConclusions=1
%%%%%%%%%%%%%%%%%%%%%%%%%%%%  Section %%%%%%%%%%%%%%%%%%%%%%%%%%%%%%%%
%
%  \section{Conclusions}
%  \label{sec:conclusions}
%  File:  conclusions.tex
%
%%%%%%%%%%%%%%%%%%%%%%%%%%%%%%%%%%%%%%%%%%%%%%%%%%%%%%%%%%%%%%%%%%%%%%

\ifnum\theOutline=1
\begin{outline}
\item DBW2 is the best
\item Careful with topology
\item Careful with finer than $2GeV$ lattice spacing
\end{outline}
\fi

In this paper we have demonstrated that the DBW2 action significantly
improves the chiral properties of domain wall fermions. The main
reason for this improvement is the suppression of gauge configurations
which support unit eigenvalues of the transfer matrix in the fifth
dimension and hence allow significant mixing of the light chiral modes
that are localized on opposite boundaries of the fifth dimension.
These problematic configurations are also those which occur as the 
topological charge of the gauge field changes during Monte Carlo evolution.
One key to improving the domain wall fermion chiral symmetry is to use
an improved gauge action which suppresses these small dislocations. This
suppression works so well in the case of the DBW2 action that at
$a^{-1}=2$ GeV and $L_s=16$ the residual chiral symmetry breaking is
roughly two orders of magnitude smaller compared to the Wilson action
case and therefore is completely negligible. Even at strong coupling
($a^{-1}\approx 1.3$ GeV) $\mres$ is about three times smaller in
physical units than for the Wilson action at 2 GeV.  In both cases
the value of $L_s$ is 16.

Besides the suppression of these small topological dislocations
associated with zero crossings of the spectral  flow of the
four-dimensional Wilson Dirac operator, we have also observed an
increased gap in the spectral flow.  Consequently, the light boundary
mode wavefunctions decay faster in the fifth dimension.  For
the DBW2 action this leads to a residual mass dependence on $L_s$
proportional to $0.6^{L_s}$. This dependence is close to Shamir's
perturbative prediction~\cite{Shamir:2000cf}.  Approaches based on the
proposals made by Shamir in~\cite{Shamir:2000cf} may be effective in
further reducing this perturbative baseline.  Work along these lines is
currently underway.

In the second part of this paper we presented results for some  
quenched hadronic observables obtained
with the DBW2 gauge action. Our conclusion is that these observables
scale very well with $a$, {\it i.e.} the good scaling of domain wall
fermions seen in quenched simulations
with the Wilson gauge action is preserved~\cite{Blum:2000kn}.

Using these improved actions, we also observed that the topological
charge of the gauge fields evolves much more slowly using standard
Monte Carlo algorithms.  In future simulations with smaller lattice
spacings, improved algorithms will be needed to efficiently sample
topology. We note, however, that this is a generic feature of all
lattice calculations which is not specific to the DBW2 action.

\fi

%%%%%%%%%%%%%%%%%%%%%%%%%%%%  Section %%%%%%%%%%%%%%%%%%%%%%%%%%%%%%%%

\ifnum\theAcknowledgments=1
\begin{acknowledgments}
The calculations reported here were done on the 400 Gflops QCDSP
computer \cite{Chen:1998cg} at Columbia University and the 600 Gflops
QCDSP computer \cite{Mawhinney:2000fx} at the RIKEN BNL Research
Center. We thank RIKEN, Brookhaven National Laboratory and the U.S.\
Department of Energy for providing the facilities essential for the
completion of this work.

 The authors would like to thank members of QCD-TARO
for providing us with their scale setting data for the DBW2 action.

This research was supported in part by the DOE under grant \#
DE-FG02-92ER40699 (Columbia), in part by the DOE under grant \#
DE-AC02-98CH10886 (Soni), in part by the RIKEN BNL Research
Center (Aoki-Blum-Dawson-Ohta-Orginos), and in part by the  
Japan Society for the Promotion of Science under a
 Fellowship for Research Abroad (Izubuchi).
\end{acknowledgments}
\fi

%%%%%%%%%%%%%%%%%%%%%%%%%%%%%% APPENDIX %%%%%%%%%%%%%%%%%%%%%%%%%%

\appendix

\ifnum\theAppendix=1
%%%%%%%%%%%%%%%%%%%%%%%%%%%%  Section %%%%%%%%%%%%%%%%%%%%%%%%%%%%%%%%
%
%  File:  appendix.tex
%
%%%%%%%%%%%%%%%%%%%%%%%%%%%%%%%%%%%%%%%%%%%%%%%%%%%%%%%%%%%%%%%%%%%%%%

\ifnum\theOutline=1
\begin{outline}
\item
\end{outline}
\fi

%%%%%%%%%%%%%%%%%%%%%%%%%%  Section  %%%%%%%%%%%%%%%%%%%%%%%%%%%%%%%%%

\section{There is no apendix}

\fi

%%%%%%%%%%%%%%%%%%%%%%%%%%%%  Bibliography %%%%%%%%%%%%%%%%%%%%%%%%%%%

\bibliography{paper}

\pagebreak

%%%%%%%%%%%%%%%%%%%%%%%%%%%%%% TABLES %%%%%%%%%%%%%%%%%%%%%%%%%%

\ifnum\theTables=1
%%%%%%%%%%%%%%%%%%%%%%%%%%%%%% TABLES %%%%%%%%%%%%%%%%%%%%%%%%%%

%%%%%%%  Mres parameters  %%%%%%%%%%%%%%%%%%
\begin{table}[t]
\caption{Simulation parameters for each gauge action tested.  The
$\rho$ mass, $m_\rho$, is given in the chiral limit for the largest
$L_s$ in each case. As usual $\beta \sim 1/g_0^2$, where $g_0$ 
is the bare gauge coupling, $M_5$ is the 
five-dimensional fermion mass, 
$L_s$ the size of the 5th dimension, $m_f$ the bare
quark mass, and $\sigma$ is the string tension computed from the heavy
quark potential.}
\label{tab:mres_sim_param}
\newcommand{\m}{\hphantom{$-$}}
\newcommand{\cc}[1]{\multicolumn{1}{c}{#1}}
\begin{ruledtabular}
\begin{tabular}{@{}lllllll}
%\hline
Action        & \cc{$\beta$}   & \cc{$M_5$} & \cc{$L_s$} & \cc{$m_f$} 
              & \cc{$m_\rho$} & \cc{$\sqrt \sigma$}\\
\hline
Wilson~\mbox{\cite{Blum:2000kn}}     
              & \m6.00  & 1.8 & 12-24 & 0.015-0.040 & 0.404(8)  &
                0.227(6)~\cite{Bali:1992ab}\\
Symanzik      & \m8.40  & 1.8 & 8-16  & 0.020-0.060 & 0.411(14) & 
                0.2278(18)~\cite{Bernard:2000gd}\\
Iwasaki       & \m2.60  & 1.8 & 16    & 0.020-0.060 & 0.415(13) & 
                 0.231(6)~\cite{Okamoto:1999hi}\\
DBW2          & \m1.04  & 1.7 & 8-16  & 0.020-0.060 & 0.399(11) & 0.2246(16)\\
%DBW2          & \m0.87  & 1.8 & 16    & 0.020-0.060 & 0.592(9)  & 0.324(6)\\
%\hline
\end{tabular}\\
\end{ruledtabular}
\end{table}

%%%%%%%%%%%%%%%%%%%%%%%%% Mres %%%%%%%%%%%%%%%%%%%%%%%% 

\begin{table}[htbp]
\caption{The residual mass $\mres$ at $a^{-1} \approx$ 2 GeV 
for the actions tested. In the construction of this table, for the Symanzik action we used 51 configurations,
for the Iwasaki 45, and for the DBW2 89.}
\label{tab:Mres}
\begin{ruledtabular}
\begin{tabular}{crrrr}
$m_f$ & $L_s$ & Symanzik & Iwasaki & DBW2 \\
\hline                  

% for the DBW2 I used the non-overrelaxed sample of lattices
% this is done for consistency amoung all values of Ls
0.020& 8&$3.04(5) 10^{-3}$&-                      &$7.54 (5) 10^{-4}$\\
0.020&12&$8.2 (4) 10^{-4}$&-                      &$9.92 (20) 10^{-5}$\\
0.020&16&$3.3 (3) 10^{-4}$&$1.4(4) 10^{-4}$&$1.60(5) 10^{-5}$\\
\hline
0.040& 8&$2.90(4) 10^{-3}$&-                     &$7.49 (5) 10^{-4}$\\
0.040&12&$7.4(3) 10^{-4}$&-                      &$9.9 (4) 10^{-5}$\\
0.040&16&$2.73(24) 10^{-4}$&$1.2(4) 10^{-4}$&$1.56(3) 10^{-5}$\\
\hline
0.060& 8&$2.82(3) 10^{-3}$&-                       &$7.50 (8) 10^{-4}$\\
0.060&12&$6.95 (23) 10^{-4}$&-                     &$1.00 (6) 10^{-4}$\\
0.060&16&$2.44 (18) 10^{-4}$&$1.15(27) 10^{-4}$&$1.565(23) 10^{-5}$\\

\end{tabular}
\end{ruledtabular}
\end{table}    

%%%%%%%%%%%%%%%%%%%%%%%%%%%%%% TABLES %%%%%%%%%%%%%%%%%%%%%%%%%%

%%%%%%%  heavy quark potential  %%%%%%%%%%%%%%%%%%
\begin{table}[htbp]
\caption{Parameters and resulting scales for the DBW2 gauge action used for
 the  spectrum
 analysis. The quantity $a_\rho$ is the lattice spacing determined by the
 $\rho$  meson mass taken from Table.~\ref{tab:rho_fit}. 
The jackknife errors are quoted for the string tension $\sigma$ and the Sommer parameter  $r_0$.}
\label{tab:DBW2param}
\begin{ruledtabular}
 \begin{tabular}{ccccccc}
  $\beta$ & $M_5$ & $L_s$ & statistics & $\sqrt{\sigma}$ & $r_0$
  & $a_\rho^{-1} \mbox{(GeV)}$ \\
  \hline
  0.87  & 1.8 & 16 & 100 & 0.324(6) & 3.58(4) & $1.31(4)$\\
  1.04  & 1.7 & 16 & 405 & 0.2246(16) & 5.24(3) & $1.98(3)$\\
 \end{tabular}
\end{ruledtabular}
\end{table}

%%%%%%%  mres  %%%%%%%%%%%%%%%%%%
\begin{table}[htbp]
\caption{The residual mass for the DBW2 gauge action calculated on the
 same configurations used to evaluate the hadronic observables. The values for
 $m_f=0$ have been obtained from a linear fit in
 $m_f$. The value for $m_f=0.09$ at $\beta=0.87$ is excluded from the fit.}
\label{tab:DBW2mres}
\newcommand{\cc}[1]{\multicolumn{1}{c}{#1}}
\begin{ruledtabular}
 \begin{tabular}{llll}
  \multicolumn{2}{c}{$\beta=0.87$} & \multicolumn{2}{c}{$\beta=1.04$}\\
\hline
  \cc{$m_f$} & \cc{$\mres$} & \cc{$m_f$} &
  \cc{$\mres$}\\
  \hline
% mres-mf.b087L16M18
% mres-mf=0.b087L16M18
% mres-mf.b104FWL16M17
% mres-mf=0.b104FWL16M17
 0.01 & 5.44 (23) $10^{-4}$ & 0.01  & 1.80  (9) $10^{-5}$ \\
 0.02 & 5.16 (22) $10^{-4}$ & 0.015 & 1.80 (11) $10^{-5}$ \\
 0.03 & 4.84 (21) $10^{-4}$ & 0.02  & 1.77 (11) $10^{-5}$ \\
 0.04 & 4.55 (19) $10^{-4}$ & 0.025 & 1.74 (11) $10^{-5}$ \\
 0.05 & 4.30 (18) $10^{-4}$ & 0.03  & 1.71 (10) $10^{-5}$ \\
 0.06 & 4.08 (16) $10^{-4}$ & 0.035 & 1.69  (8) $10^{-5}$ \\
 0.09 & 3.52 (13) $10^{-4}$ & 0.04  & 1.67  (7) $10^{-5}$ \\
 0    & 5.69 (26) $10^{-4}$ & 0     & 1.86 (11) $10^{-5}$ \\
 \end{tabular}
\end{ruledtabular}
\end{table}

%%%%%%% raw data of masses %%%%%%%%%%%%%%%%%%%%%%%% 
\begin{table}[htbp]
\caption{Hadron masses computed using the DBW2 gauge action. 
The superscripts PP and AA refer to the 
pseudoscalar and axial vector correlation functions, respectively.
All masses are obtained with degenerate quarks.}
\label{tab:hadronmass}
\newcommand{\cc}[1]{\multicolumn{1}{c}{#1}}
\begin{ruledtabular}
\begin{tabular}{llllll}
\cc{$\beta$} & \cc{$m_f$} & \cc{$m_\pi^{PP}$} & \cc{$m_\pi^{AA}$}
 & \cc{$m_\rho$} & \cc{$m_N$} \\
\hline
% 1/mass.b087L16M18.RHO-mf
% 1/mass.b087L16M18.NUC0-mf
% 1/mass.b087L16M18.PION-mf
% 1/mass.b087L16M18.AA-mf
 & 0.01 & 0.2248 (25) & 0.2179 (31) & 0.607  (22) & 0.790  (75) \\
 & 0.02 & 0.2997 (19) & 0.2966 (23) & 0.640  (17) & 0.871  (23) \\
 & 0.03 & 0.3603 (16) & 0.3590 (20) & 0.662  (11) & 0.921  (14) \\
0.87
 & 0.04 & 0.4128 (15) & 0.4118 (19) & 0.685   (8) & 0.975  (10) \\
 & 0.05 & 0.4601 (14) & 0.4589 (18) & 0.709   (6) & 1.021   (8) \\
 & 0.06 & 0.5037 (13) & 0.5021 (17) & 0.732   (5) & 1.067   (7) \\
 & 0.09 & 0.6192 (13) & 0.6170 (15) & 0.803   (4) & 1.197   (6) \\
\hline
% 2/mass.b104FWL16M17.RHO-mf
% 2/mass.b104FWL16M17.NUC0-mf
% 2/mass.b104FWL16M17.PION-mf
% 2/mass.b104FWL16M17.AA-mf
 & 0.01  & 0.1794 (22) & 0.1759 (21) & 0.413 (6)   & 0.546 (13) \\
 & 0.015 & 0.2098 (17) & 0.2075 (18) & 0.424 (4)   & 0.575 (9)  \\
 & 0.02  & 0.2377 (15) & 0.2359 (16) & 0.435 (4)   & 0.602 (7)  \\
1.04							    
 & 0.025 & 0.2631 (13) & 0.2617 (15) & 0.447 (3)   & 0.628 (6) \\
 & 0.03  & 0.2868 (12) & 0.2857 (14) & 0.4586 (29) & 0.652 (5) \\
 & 0.035 & 0.3090 (12) & 0.3081 (13) & 0.4705 (26) & 0.676 (4) \\
 & 0.04  & 0.3300 (11) & 0.3293 (12) & 0.4825 (24) & 0.699 (4) \\
\end{tabular}
\end{ruledtabular}
\end{table}

%%%%%%% linear fit of pi %%%%%%%%%%%%%%%%%%%%%%%% 
\begin{table}[htbp]
\caption{Results from fitting $m_\pi^2$ to the linear function
 $c_0 + c_1 m_f$. The column $m_f$ shows the fitting range.}
\label{tab:pi-linearfit}
\newcommand{\cc}[1]{\multicolumn{1}{c}{#1}}
\begin{ruledtabular}
\begin{tabular}{ccccclc}
 $\beta$ & correlator & $m_f$ & $c_0$ & $c_1$ & \cc{$\chi^2$}
 & $dof$ \\
 \hline
 0.87  & PP & 0.01--0.06 & 0.0090 (14) & 4.057 (29) & 2.8 (1.2) & 4\\
 0.87  & AA & 0.01--0.06 & 0.0063 (16) & 4.090 (37) & 0.17 (35) & 4\\
 1.04  & PP & 0.01--0.04 & 0.0056 (9) & 2.566 (21) & 3.4 (8) & 5\\
 1.04  & PP & 0.015--0.04 & 0.0047 (9) & 2.597 (19) & 1.10 (25) & 4\\
 1.04  & AA & 0.01--0.04 & 0.0044 (9) & 2.585 (21) & 2.6 (7) & 5\\
 1.04  & AA & 0.015--0.04 & 0.0035 (9) & 2.615 (20) & 0.86 (23) & 4\\
\end{tabular} 
\end{ruledtabular}
\end{table}

%%%%%%% chiral log fit of pi %%%%%%%%%%%%%%%%%%%%%%%% 
\begin{table}[htbp]
\caption{Fit results for the 
 quenched chiral log contribution to the pion mass determined from the
 AA correlator using eq.~\ref{eq:chilog}.
 The symbol ``-'' indicates
 that the parameter is  constrained to be zero.}
\label{tab:chilog}
\begin{ruledtabular}
\begin{tabular}{ccccccc}
 $\beta$ & $m_f$ & $a_\pi$ & $b$ & $\delta$ & $\chi^2$ & $dof$ \\
 \hline
% cfjk.1/cfcl1b087L16M18.AA
% cfjk.1/cfcl2b087L16M18.AA
 0.87 & 0.01-0.06 &  4.04 (5) & - & 0.031 (14) & 1.6 (1.5) & 4\\
 0.87 & 0.01-0.09 &  3.40 (23) & 6.4(1.6) & 0.107 (38) & 0.19 (25) & 4\\
 1.04 & 0.015-0.04 &  2.583 (28) & - & 0.049 (14) & 2.0 (5) & 4\\
\end{tabular}
\end{ruledtabular}
\end{table}

%%%%%%% linear fit of rho %%%%%%%%%%%%%%%%%%%%%%%% 
\begin{table}[htbp]
\caption{Results from fitting the vector meson mass to a 
 linear function, $c_0 + c_1 m_f$.
 Values at the physical points and the J parameter are also listed.}
\label{tab:rho_fit}
\newcommand{\cc}[1]{\multicolumn{1}{c}{#1}}
\begin{ruledtabular}
\begin{tabular}{cclccclll}
  $\beta$ & $m_f$ & \multicolumn{1}{c}{$c_0$} & $c_1$ & $\chi^2$ & $dof$
  & \cc{$m_\rho$} & \cc{$m_{K^*}/m_\rho$} & \cc{J}\\
 \hline
 0.87  & 0.01--0.06 & 0.590 (19) & 2.37 (28) & 0.14 (50) & 4
   & 0.589 (19) & 1.138 (11) & 0.377 (37)\\
 1.04  & 0.01--0.04 & 0.389 (6) & 2.34 (12) & 0.08 (20) & 5
   & 0.388 (6) & 1.136 (5) & 0.387 (16)\\
\end{tabular}
\end{ruledtabular}
\end{table}

%%%%%%% linear fit of nucleon %%%%%%%%%%%%%%%%%%%%%%%% 
\begin{table}[htbp]
\caption{Results from fitting the baryon mass to a linear function, 
 $c_0 + c_1 m_f$.
 Values at the physical point are also listed.}
\label{tab:nuc_fit}
\begin{ruledtabular}
\begin{tabular}{ccccccc}
 $\beta$ & $m_f$ & $c_0$ & $c_1$ & $\chi^2$ & $dof$ & $m_N$ \\
 \hline
 0.87  & 0.01--0.06 & 0.783 (27) & 4.76 (44) & 0.5 (1.6) & 4 & 0.780 (27) \\
 1.04  & 0.01--0.04 & 0.502 (12) & 4.96 (26) & 0.5 (8) & 5 & 0.502 (12) \\
\end{tabular}
\end{ruledtabular}
\end{table}

%%%%%%% f_PS %%%%%%%%%%%%%%%%%%%%%%%% 
\begin{table}[htbp]
\caption{The axial current renormalization factor and pseudoscalar decay
 constants.}
\label{tab:fps-mf}
\begin{ruledtabular}
\begin{tabular}{clccc}
 $\beta$ & \multicolumn{1}{c}{$m_f$} & $Z_A$ & $f_{PS}^{AA}$ & $f_{PS}^{PP}$ \\
 \hline
% za-mf.b087PL16M18
% 1/fps1s.b087PL16M18.AA-mf
% 1/fps1s.b087PL16M18.PION-mf
      & 0.02 & 0.78199 (37) & 0.1084 (31) & 0.1078 (34) \\
      & 0.03 & 0.78404 (31) & 0.1121 (27) & 0.1110 (27) \\
 0.87 & 0.04 & 0.78612 (28) & 0.1165 (26) & 0.1145 (24) \\
      & 0.05 & 0.78824 (26) & 0.1208 (26) & 0.1192 (24) \\
      & 0.06 & 0.79042 (25) & 0.1247 (26) & 0.1237 (24) \\
 \hline
% za-mf.b104FWL16M17 
% 2/fps1s.b104FPL16M17.AA-mf
% 2/fps1s.b104FPL16M17.PION-mf
      & 0.01  & 0.84142 (17) & 0.0703 (22) & 0.0697 (21) \\
      & 0.015 & 0.84191 (14) & 0.0716 (17) & 0.0720 (17) \\
      & 0.02  & 0.84244 (12) & 0.0732 (14) & 0.0743 (15) \\
 1.04 & 0.025 & 0.84300 (11) & 0.0751 (12) & 0.0767 (14) \\
      & 0.03  & 0.84358 (10) & 0.0771 (11) & 0.0790 (13) \\
      & 0.035 & 0.84417  (9) & 0.0790 (11) & 0.0813 (12) \\
      & 0.04  & 0.84478  (9) & 0.0809 (10) & 0.0835 (12) \\
\end{tabular}
\end{ruledtabular}
\end{table}

%%%%%%% f_PS @ physical point %%%%%%%%%%%%%%%%%%%%%%%% 
\begin{table}[htbp]
\caption{The axial current renormalization factor at $m_f=-\mres$ and 
 the pseudoscalar decay constants at physical points in [MeV].}
\label{tab:fps}
\begin{ruledtabular}
\begin{tabular}{cccccccc}
% cfza.b087PL16M18
% cfza.b104WL16M17
 $\beta$ & $Z_A(m_f=-\mres)$ & $f_{\pi}^{AA} $ & $f_{\pi}^{PP}$ 
 & $f_{K}^{AA} $ & $f_{K}^{PP}$ & $f_{K}^{AA}/f_{\pi}^{AA}$ 
 & $f_{K}^{PP}/f_{\pi}^{PP}$ \\
 \hline
 0.87 & 0.77759(45) & 130.4(6.7) & 129.0(7.3) & 148.9(5.2) & 147.3(5.4)
   & 1.142(26) & 1.141(30)\\
 1.04 & 0.84018(18) & 130.8(4.9) & 129.0(5.0) & 147.4(3.3) & 149.7(3.6)
   & 1.139(24) & 1.118(25)\\
\end{tabular}
\end{ruledtabular}
\end{table}

\clearpage
\pagebreak

\fi
\pagebreak

%%%%%%%%%%%%%%%%%%%%%%%%%%%%%% FIGURES %%%%%%%%%%%%%%%%%%%%%%%%%%%%%%%%

\ifnum\theFigures=1
%%%%%%%%%%%%%%%%%%%%%%%%%%%%%%%%%%%%%%%%% 
%
%  Figures 
%

%%%%%%%%%%  Constancy of chiral log over our mass range %%%%%%%%%%%%%%
%
%
%

\begin{figure}
\begin{center}
\includegraphics[width=\textwidth]{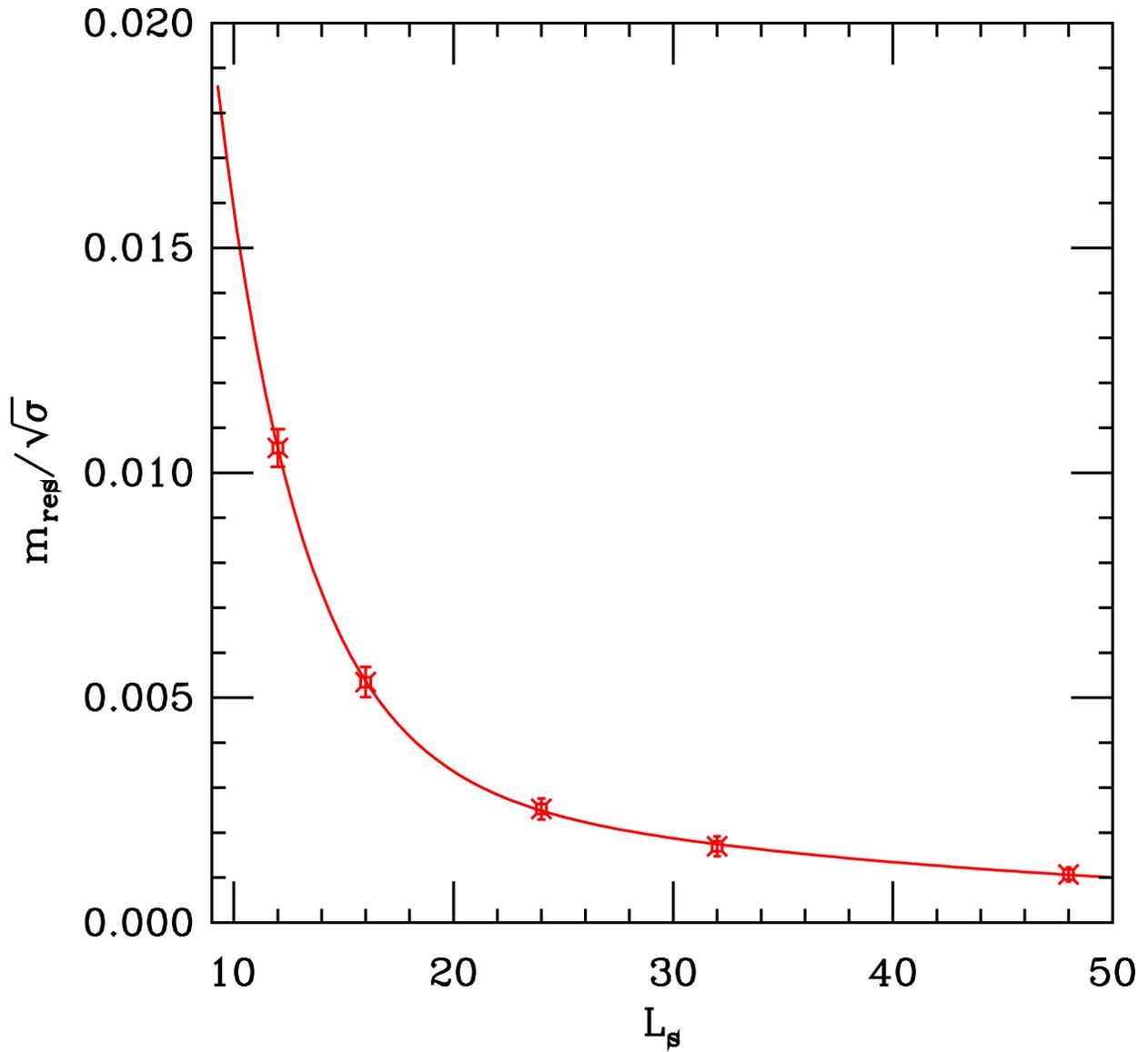}
\end{center}
\caption{The residual mass normalized by the square root of the string tension
($\sigma$) as function of the $L_s$  for
the Wilson gauge action at $\beta=6.0$ (results are from~\cite{Blum:2000kn}).
The solid line is a fit to a double exponential. At large $L_s$ the
decay is rather weak.}
\label{fig:wilson_mres}
\end{figure}

\begin{figure}
\begin{center}
\includegraphics[width=\textwidth]{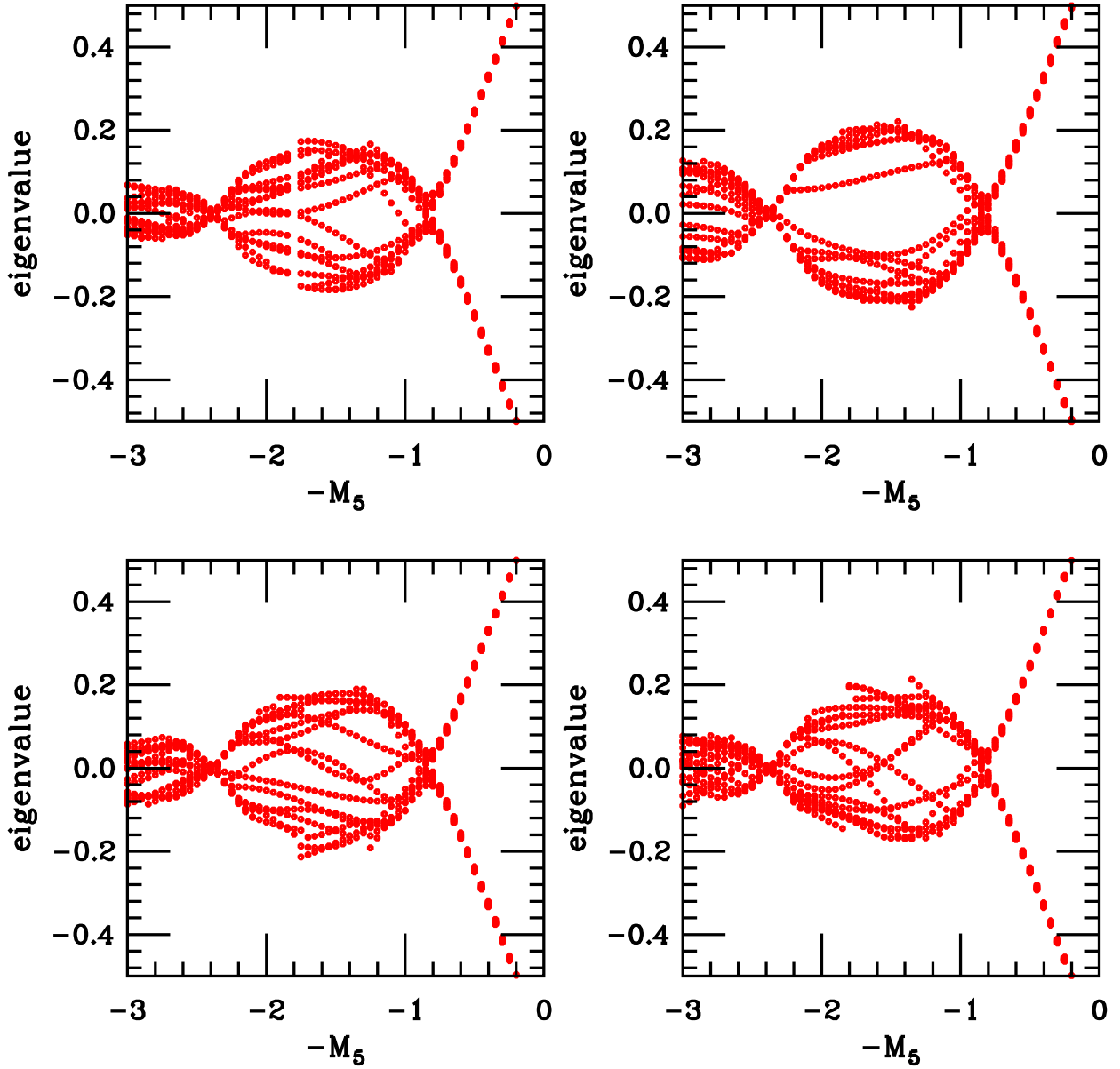}
\end{center}
\caption{Spectral flows of the Hermitian Wilson Dirac operator
 $\gamma_5 D_W$ for
 four typical gauge configurations generated with
 the Wilson gauge action at $\beta=6.0$. There are many crossings in the
neighborhood of $M_5=1.8$ which
 induce explicit chiral symmetry breaking for domain wall
 fermions. The size of the would-be gap in the region of the 
 five-dimensional fermion mass $M_5\approx 1.8$ is also relatively small
 compared to the obvious gap above the critical Wilson mass
 $M_5\approx 0.8$.
 Both effects enhance mixing of the light domain wall fermion modes and
 hence the value of the $m_{res}$.}
\label{fig:wilson_flow}
\end{figure}

\begin{figure}
\begin{center}
\includegraphics[width=\textwidth]{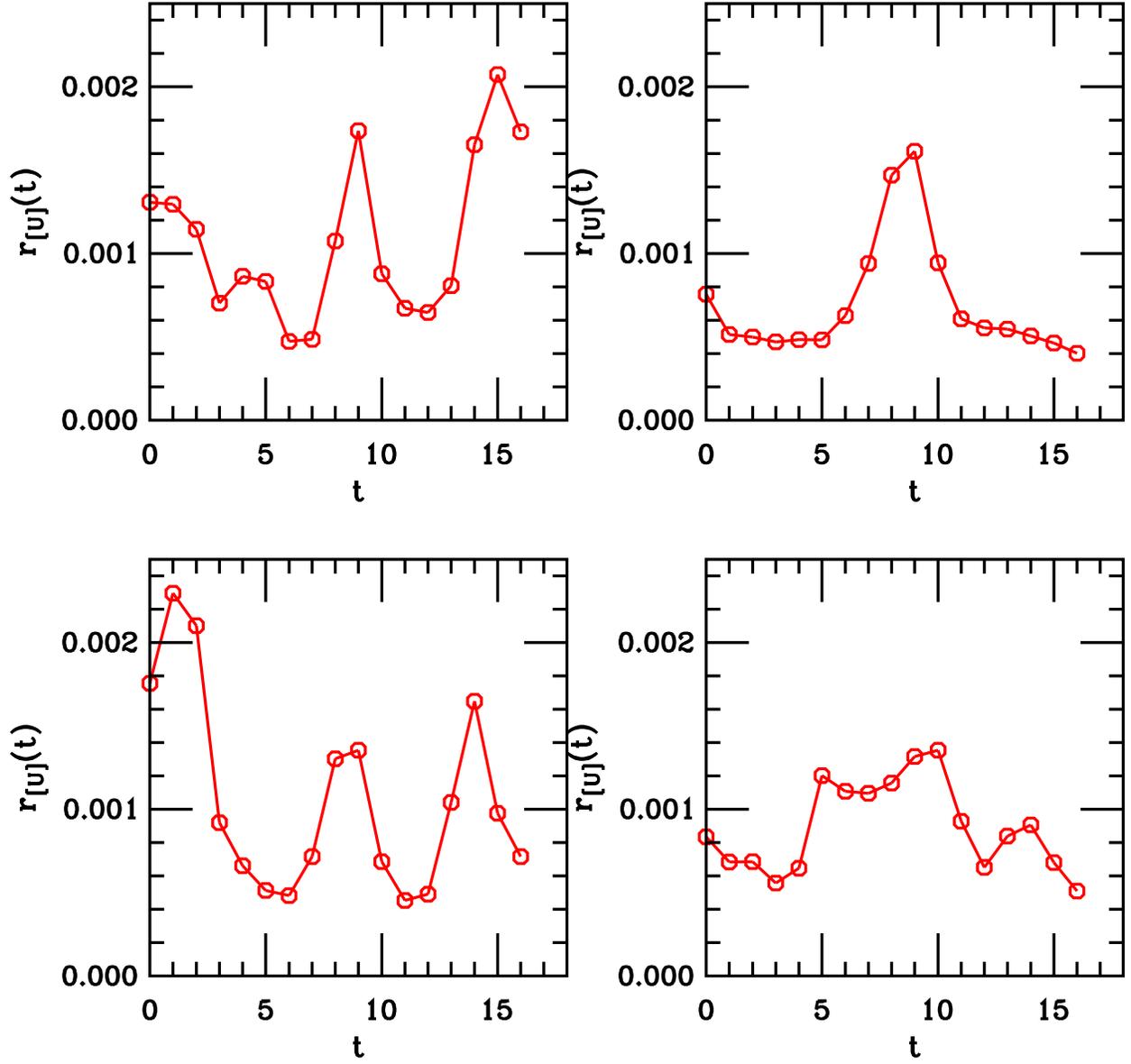}
\end{center}
\caption{
The ratio of Eq.~\ref{eq:cnfMres} for four Wilson gauge configurations.
The bare quark mass is $0.02$ and $M_5=1.8$. The configurations
used are the same as those in Fig.~\ref{fig:wilson_flow} and
the panels are in one to one correspondence with the panels in  
Fig.~\ref{fig:wilson_flow}}
\label{fig:four_wilson_spikes}
\end{figure}

\clearpage

\begin{figure}
\begin{center}
\includegraphics[width=\textwidth]{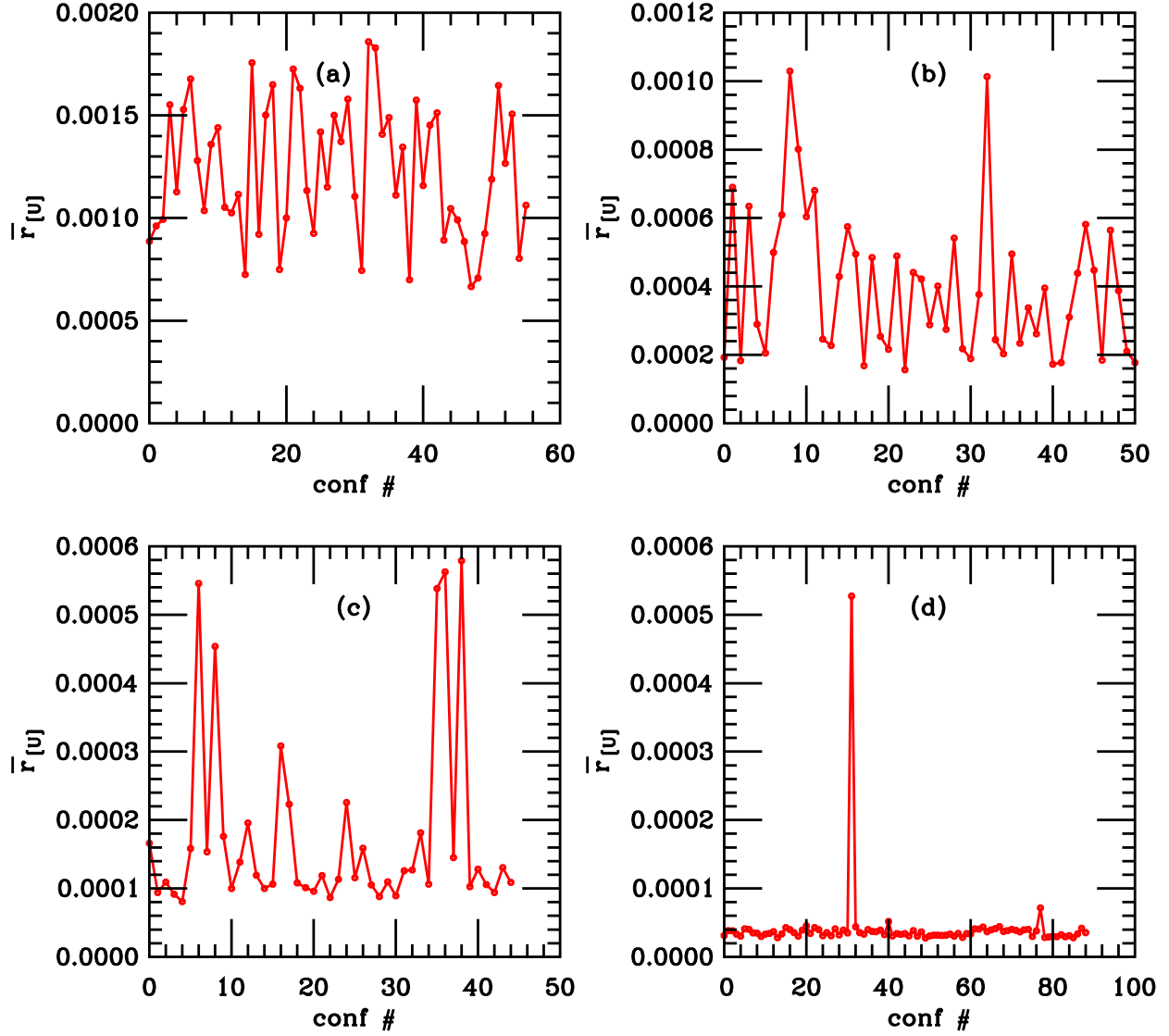}
\end{center}
\caption{The quantity $\bar{r}_{[U]}$ defined in
 Eq.~\ref{eq:sum_R}  vs. configuration number for
(a) the Wilson gauge action at $\beta=6.0$, 
(b) the Symanzik gauge action at $\beta=8.4$,
(c) the Iwasaki gauge action at $\beta=2.6$ ,
(d) and the DBW2 gauge action at $\beta=1.04$. All four cases
correspond to $a^{-1} \approx 2\; {\rm GeV}$. The large spikes seen
in the Wilson case are significantly reduced for the Iwasaki action
and almost eliminated for the DBW2 action. These spikes corresponded to
zero eigenvectors of $\gamma_5 D_W$ and are a
significant source of chiral symmetry breaking for domain wall fermions.}
\label{fig:spikes}
\end{figure}

\begin{figure}
\begin{center}
\includegraphics[width=\textwidth]{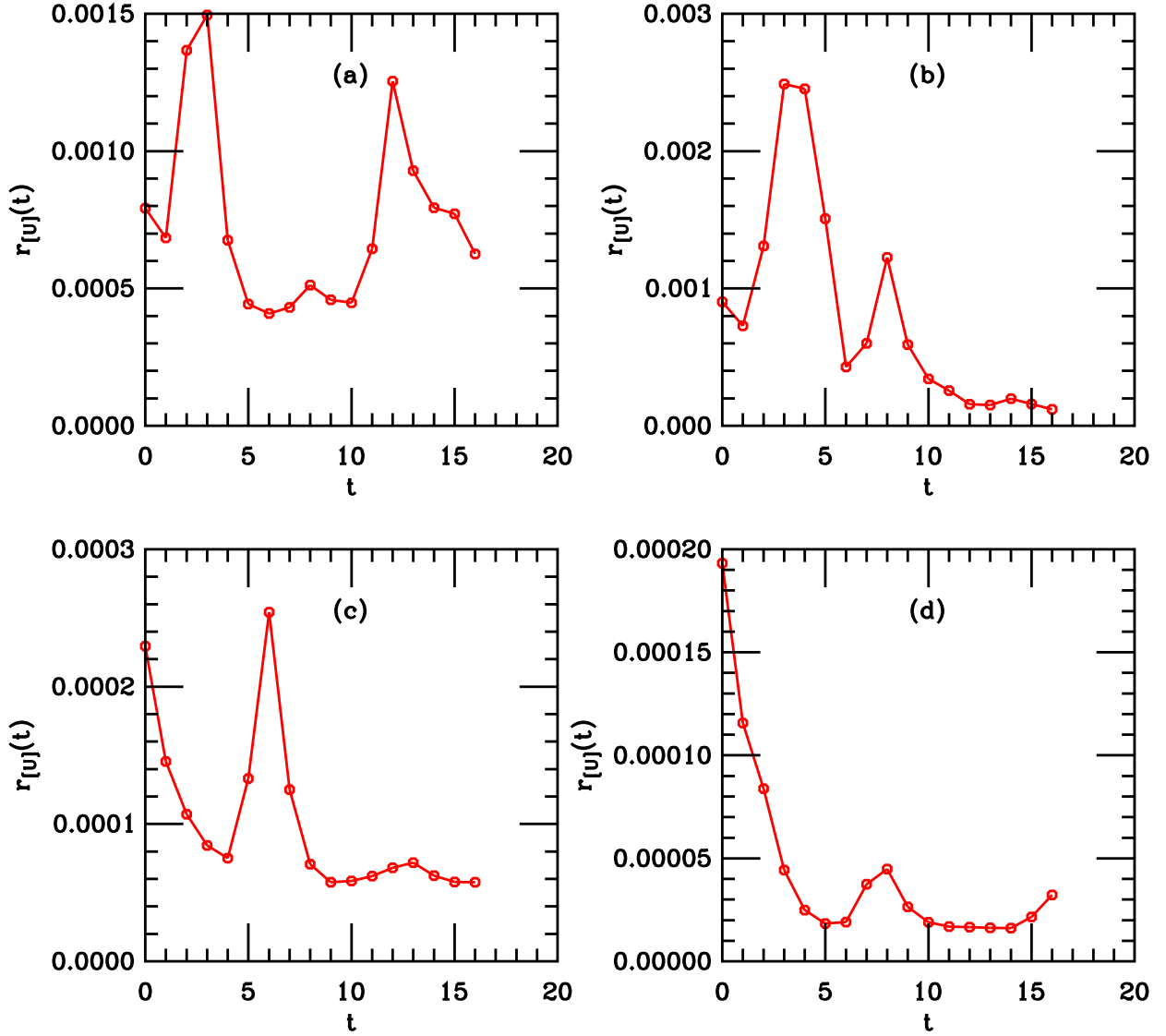}
\end{center}
\caption{The ratio of Eq.~\ref{eq:cnfMres} for a single configuration  of 
(a) the Wilson gauge action at $\beta=6.0$,
(b) the Symanzik gauge action at $\beta=8.4$,
(c) the Iwasaki gauge action at $\beta=2.6$,
(d) and the DBW2 gauge action at $\beta=1.04$.
The spikes are quite localized in Euclidean time $t$. 
Examination of the eigenvectors of the
(domain wall fermion or Wilson) Dirac operator confirms the zero modes
are localized in space as well. 
}
\label{fig:one_spike}
\end{figure}

\begin{figure}
\begin{center}
\includegraphics[width=\textwidth]{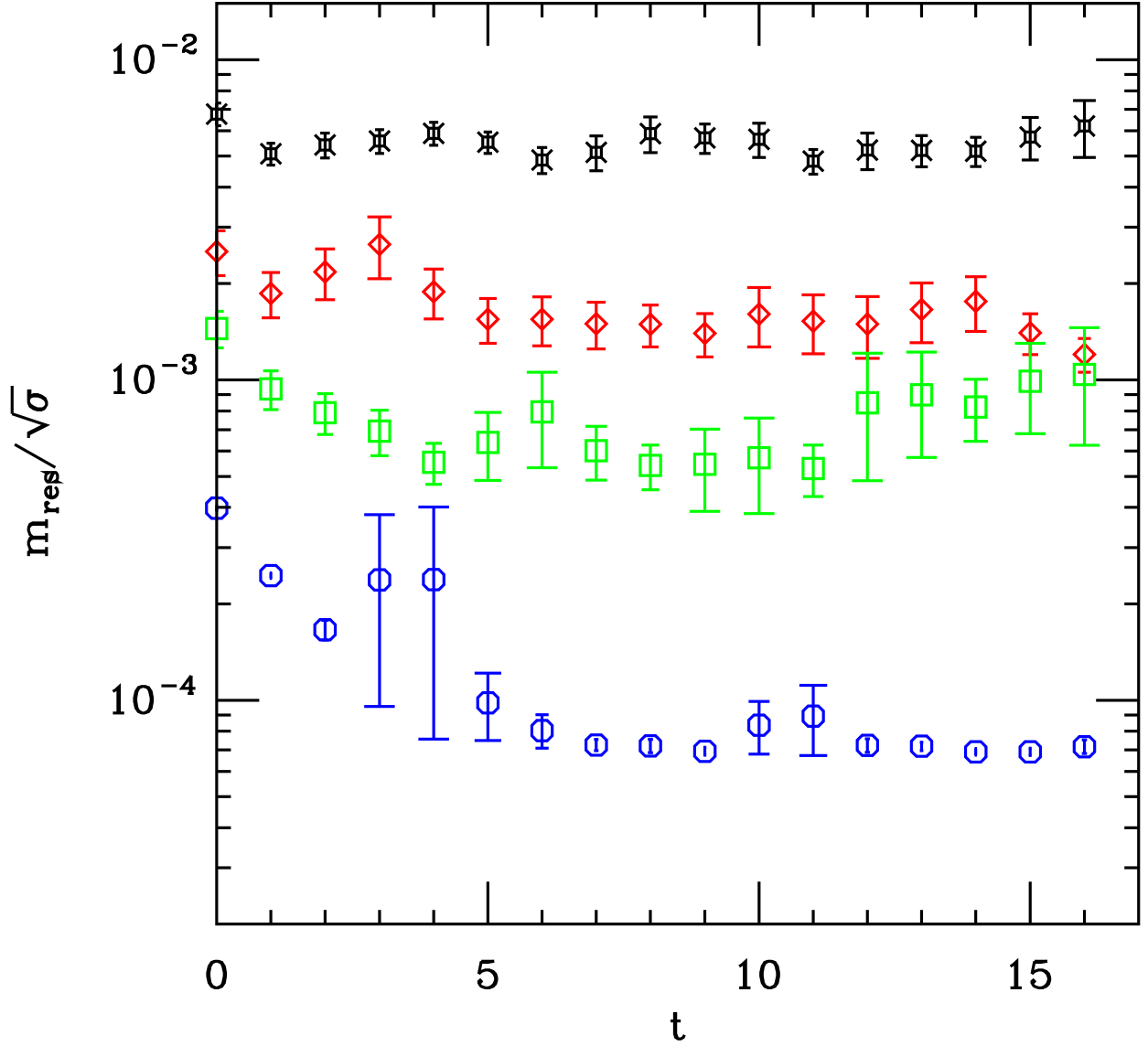}
\end{center}
\caption{The ratio defined in Eq.~\ref{eq:Mres_t} at $a^{-1}\approx 2$ GeV.  
The  fancy squares  correspond to the Wilson gauge action, 
the diamonds to Symanzik, the squares to Iwasaki,
and the octagons to DBW2. The bare quark mass in all cases is 0.020
 and $L_s=16$.}
\label{fig:mres_vs_t}
\end{figure}

\begin{figure}
\begin{center}
\includegraphics[width=\textwidth]{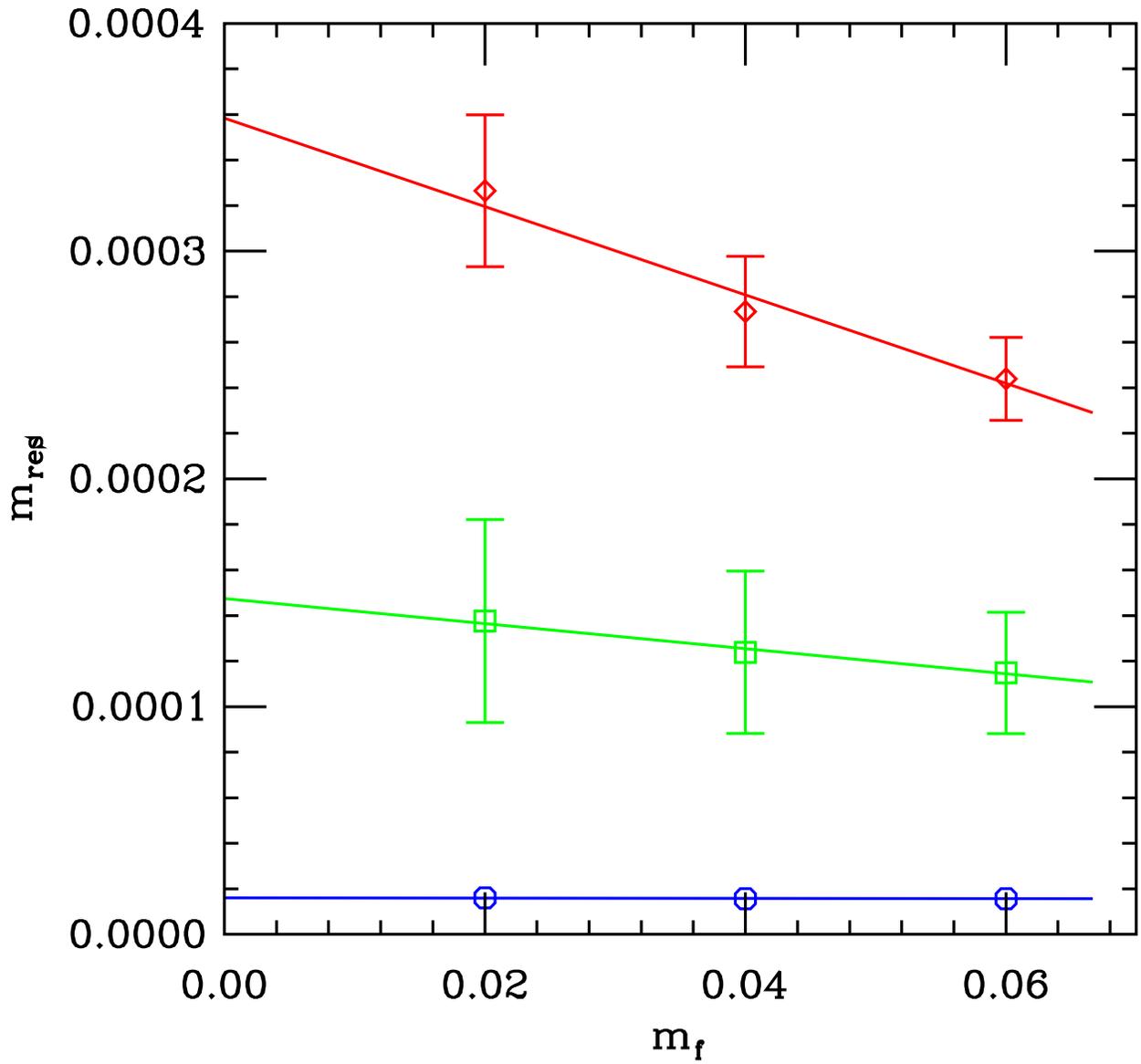}
\end{center}
\caption{The residual mass at $a^{-1}\approx 2$ GeV 
as a function of the bare quark mass.  
The octagons correspond to DBW2, the squares to Iwasaki, and  
the diamonds to Symanzik. In each case $L_s=16$.}
\label{fig:mres_vs_mq}
\end{figure}

\clearpage

\begin{figure}
\begin{center}
\includegraphics[width=\textwidth]{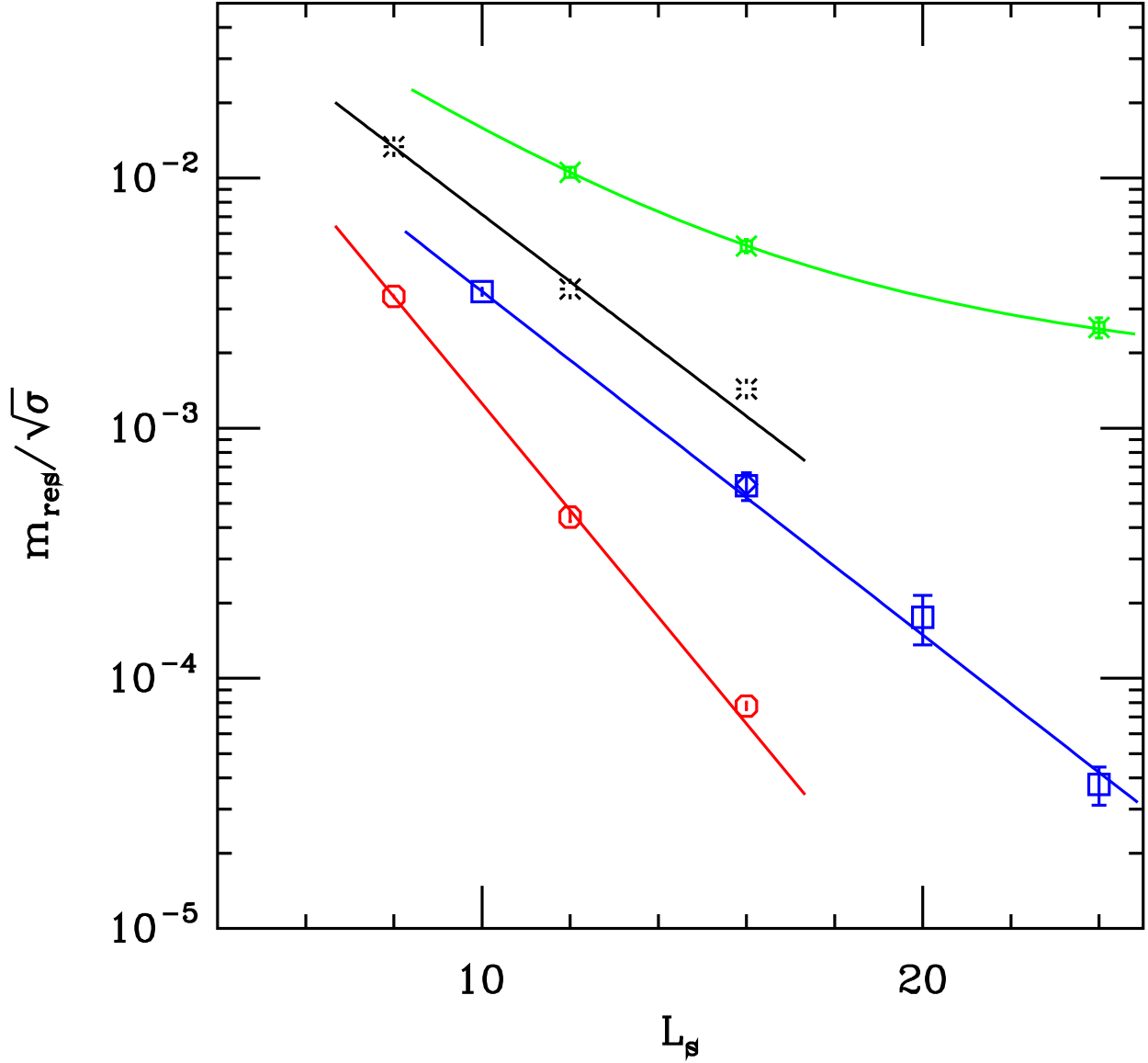}
\end{center}
\caption{Dependence of the residual mass on the size of the 5th dimension
at $a^{-1}\approx 2$ GeV. The
octagons correspond to DBW2, the squares
(CP-PACS~\cite{AliKhan:2000iv}) and diamond (RBC~\cite{Blum:2000kn}) 
to Iwasaki, the bursts to Symanzik, and the fancy squares to Wilson.
All but the Wilson action 
fit a simple exponential decay reasonably well. Note the Iwasaki
results use different gauge field ensembles at each value of $L_s$.
In the case of the Wilson action, the results are fit to a double
exponential function.
}
\label{fig:mres_vs_ls}
\end{figure}

\clearpage

\begin{figure}
\begin{center}
\includegraphics[width=\textwidth]{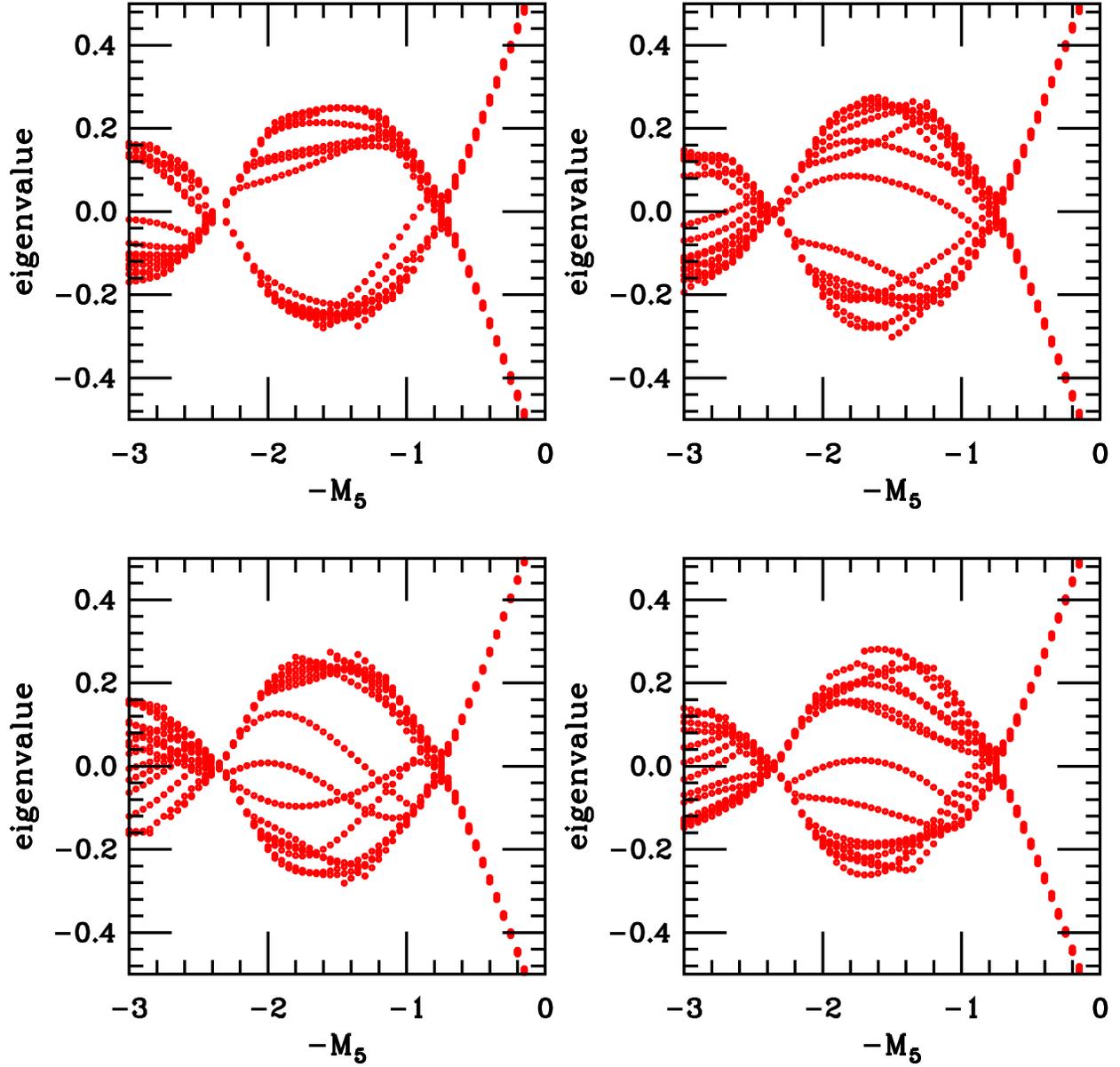}
\end{center}
\caption{The same as Fig.~\ref{fig:wilson_flow}, but 
for the Symanzik gauge action at $\beta=8.4$. The number of crossings in the
neighborhood of $M_5=1.8$
appears slightly smaller and the would-be gap slightly larger than in
the Wilson case.} 
\label{fig:Symanzik_flow}
\end{figure}

\begin{figure}
\begin{center}
\includegraphics[width=\textwidth]{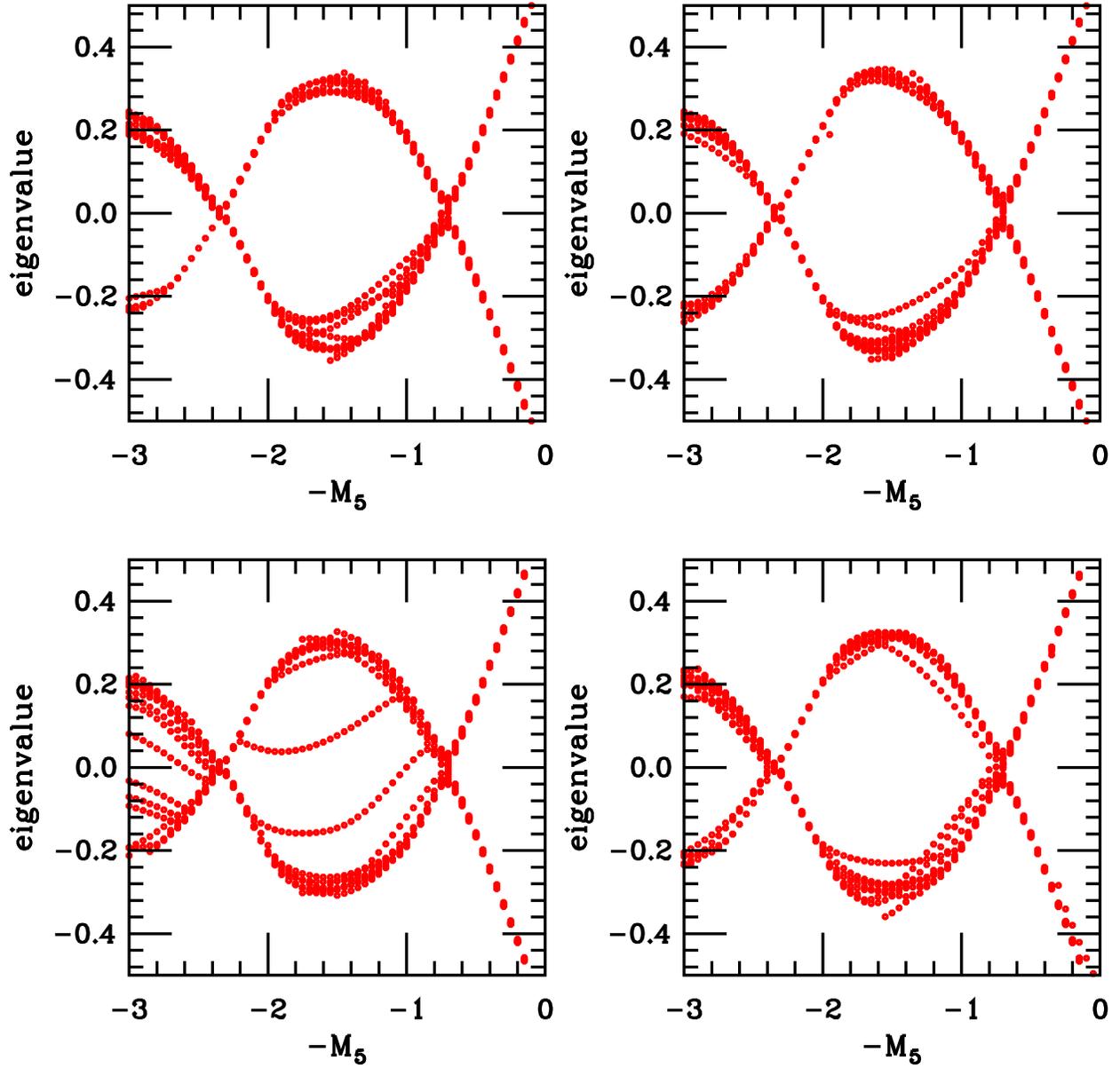}
\end{center}
\caption{The same as Fig.~\ref{fig:wilson_flow}, but 
for the Iwasaki gauge action at $\beta=2.6$. The number of crossings
neighborhood of $M_5=1.8$
is significantly smaller and the gap clearly larger than in
the Wilson case.
}
\label{fig:iwasaki_flow}
\end{figure}

\begin{figure}
\begin{center}
\includegraphics[width=\textwidth]{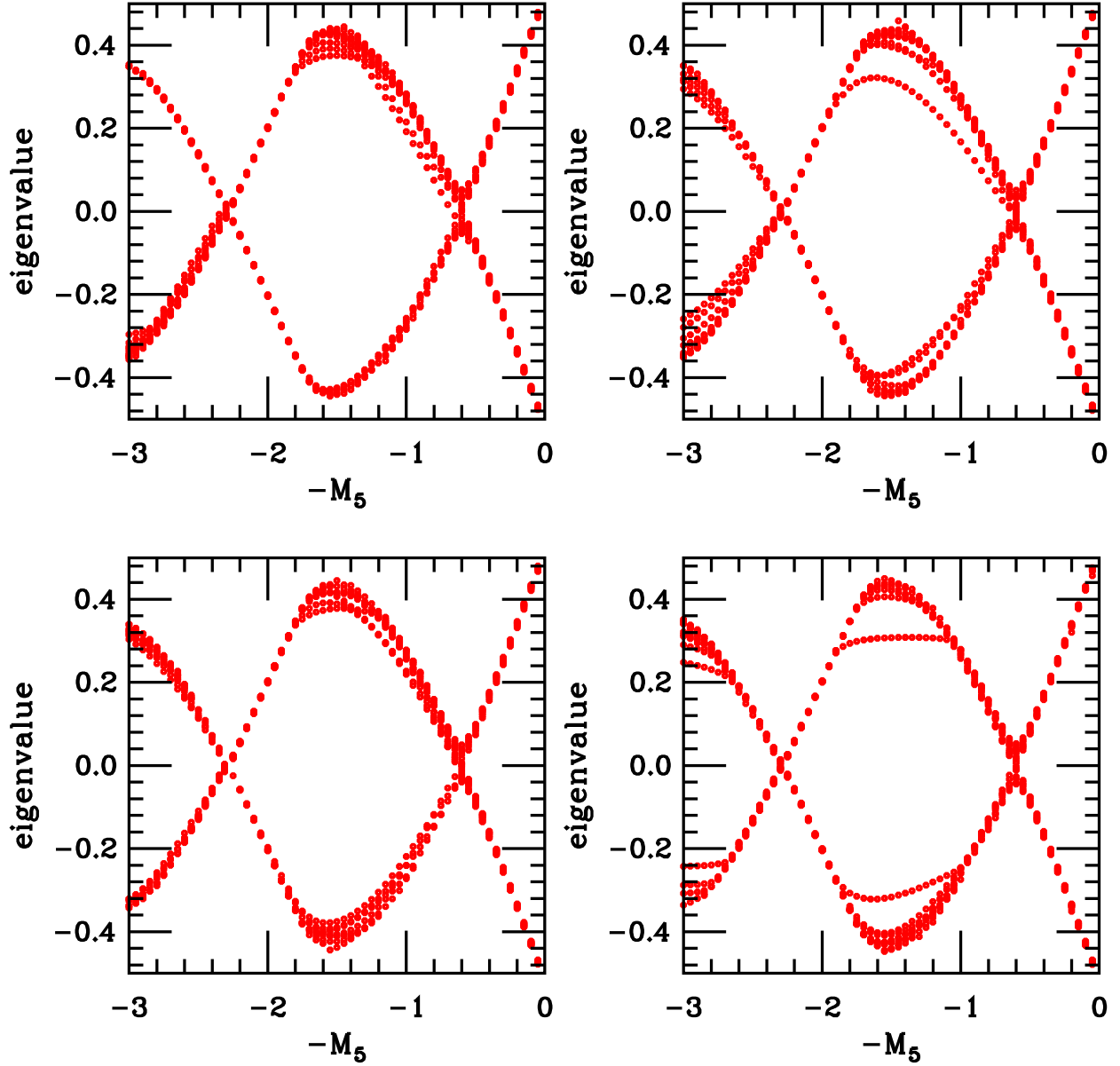}
\end{center}
\caption{The same as Fig.~\ref{fig:wilson_flow}, but 
for the DBW2 gauge action at $\beta=1.04$. There are no crossings
neighborhood of $M_5=1.8$
and the gap is quite large, roughly comparable to the gap at the
corresponding mass above the critical Wilson mass.
Note that this is also true for the region beyond the next critical 
Wilson mass, $M_5\approx 2.3$, where the four flavor Wilson fermion doublers 
become light.}
\label{fig:dbw2_flow}
\end{figure}

\newpage

\begin{figure}
\begin{center}
\includegraphics[width=\textwidth]{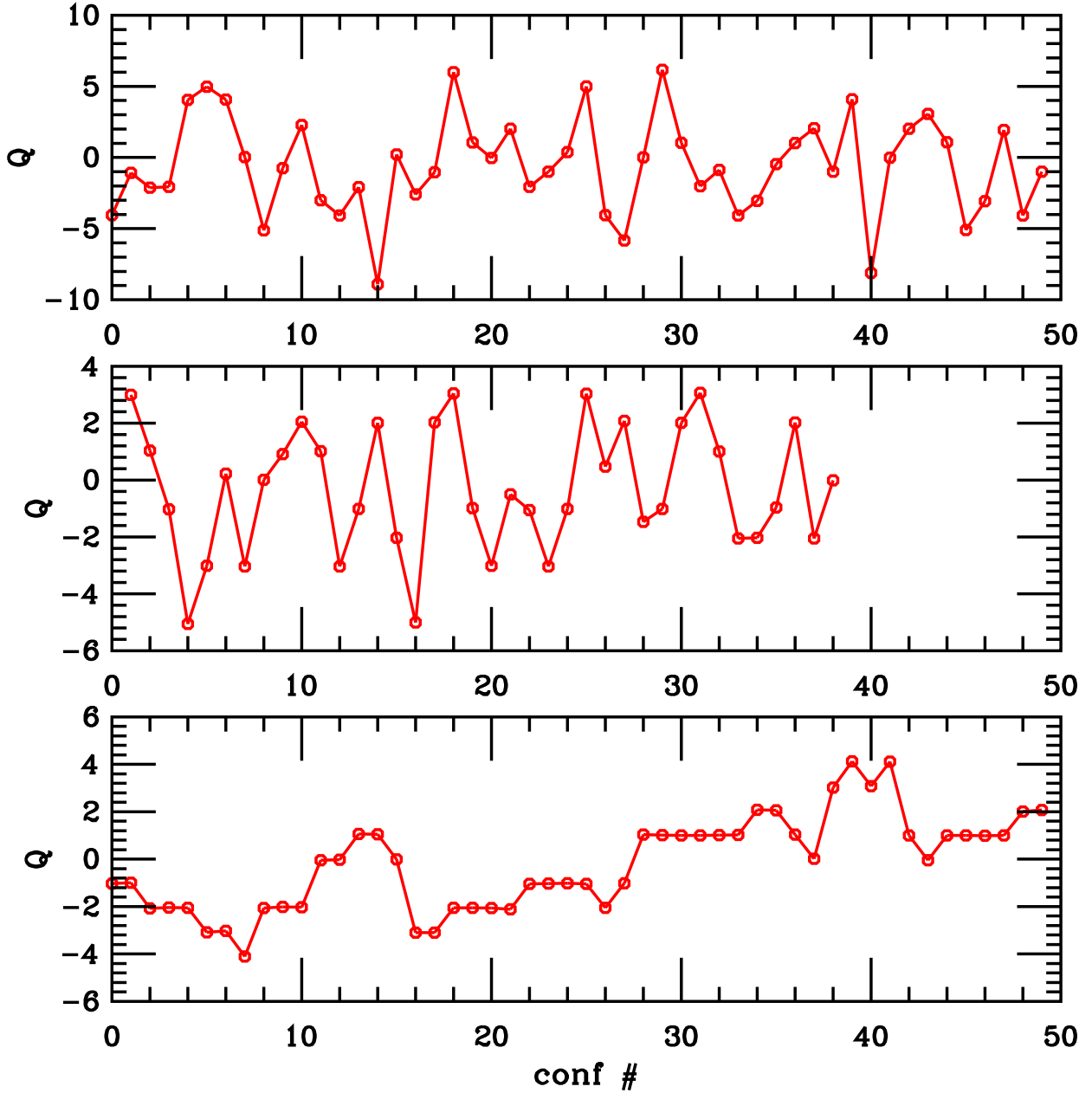}
\end{center}
\caption{The configuration 
history of the topological charge $Q$. The configurations are separated by
1000 sweeps of Cabibo-Marinari pseudo-heatbath 
with a Kennedy-Pendleton accept/reject step. At the top is the Symanzik
action with lattice size $16^3\times 32$ and coupling $\beta=8.4$. 
In the middle is the Iwasaki action with size 
$16^4$ and $\beta=2.6$. At the bottom
is the DBW2 action with size $16^3\times 32$, and
$\beta=1.04$. All cases correspond to roughly the same scale,
$a^{-1}\approx 2$ GeV. The DWB2 action, which suppresses
configurations with small instantons, shows a significant reduction of the
tunneling between topological charge sectors.} 
\label{fig:qtop}
\end{figure}

\clearpage

%%%%%%%%%%%%%%%%%%%%%%%%%%%%%%%%%%%%%%%%% 
%
%  Figures 
%

%%%%%%%%%%  Constancy of chiral log over our mass range %%%%%%%%%%%%%%
%
%
%

\begin{figure}
\begin{center}
\includegraphics[width=11cm]{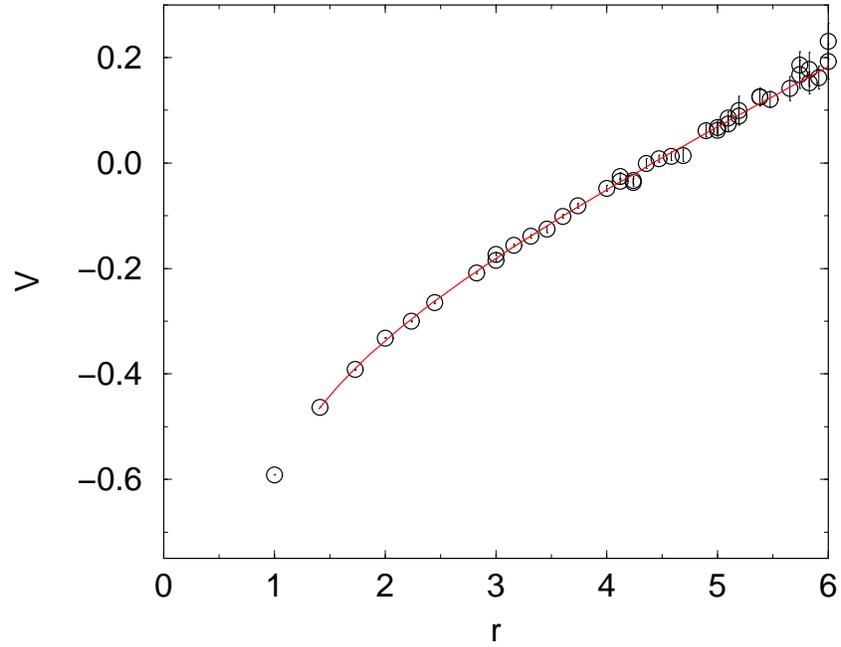}
\end{center}
\caption{The heavy quark potential for DBW2, $\beta=0.87$. The solid line
denotes the fit to Eq.~\ref{eq:potential_fit} 
from which the string tension is determined.}
\label{fig:potb087}
\end{figure}

\begin{figure}
\begin{center}
\includegraphics[width=11cm]{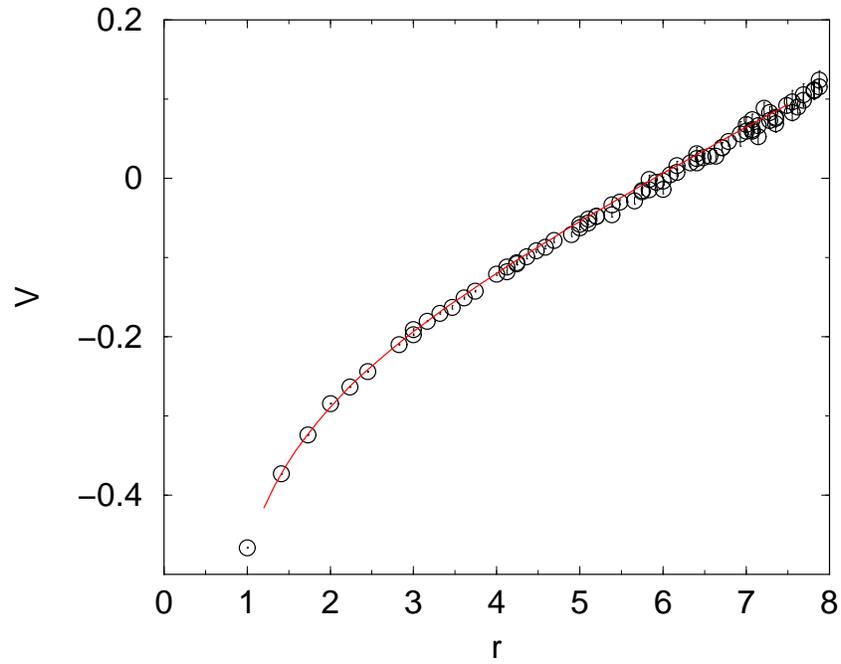}
\end{center}
\caption{The same as Fig.~\ref{fig:potb087} but for $\beta=1.04$.}
\label{fig:potb104}
\end{figure}

\begin{figure}
\begin{center}
\includegraphics[width=12cm]{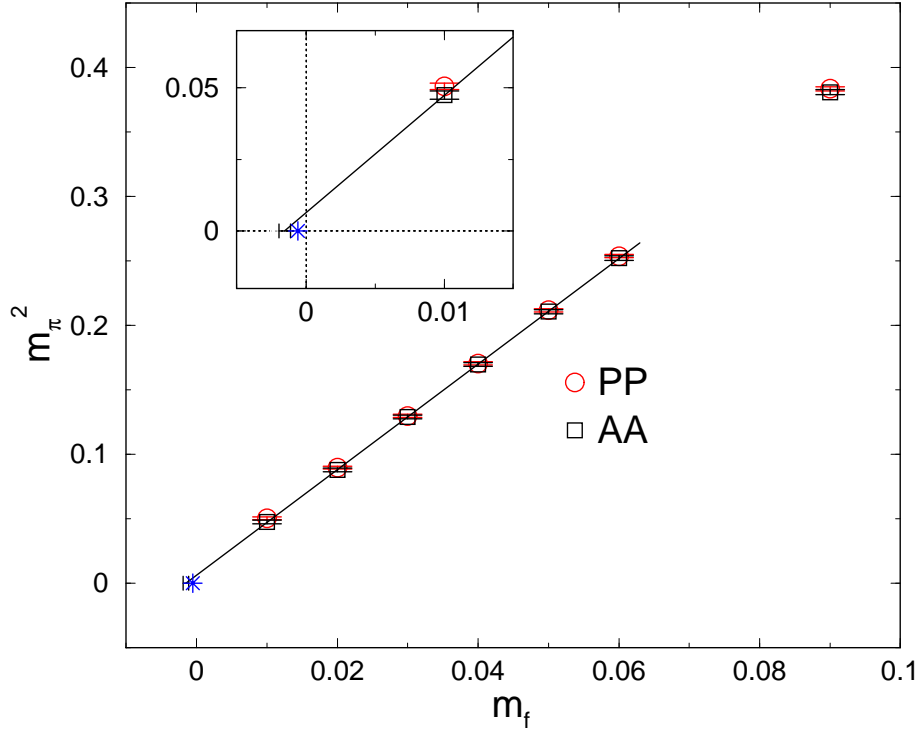}
\end{center}
\caption{The quenched pion mass squared as a function of $m_f$ for
$\beta=0.87$. Masses extracted from the pseudoscalar (PP) and axial vector
(AA) correlation functions are shown. They agree quite well, except
for the point at $m_f=0.01$ where there is a small difference outside of
statistical errors. 
 The physical volume is roughly $(2.4\,\,{\rm fm})^3$, so
contamination from topological
 zero modes which can induce such splittings (see text) is suppressed.
 The line is a fit to a simple linear function. The
 extrapolation slightly overshoots the expected chiral limit point $m_f
= -\mres$ (see inset), suggesting a quenched chiral log.}
\label{fig:pi-mf_b087}
\end{figure}

\begin{figure}
\begin{center}
\includegraphics[width=12cm]{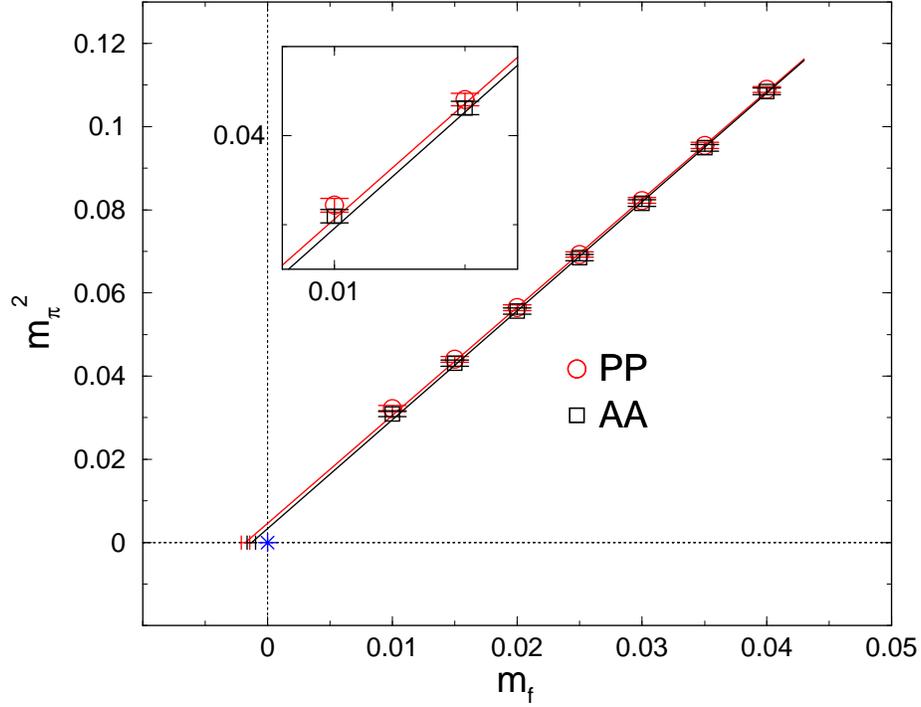}
\end{center}
\caption{The same as Fig.~\ref{fig:pi-mf_b087}, but for $\beta=1.04$. Here
the physical volume is roughly $(1.6\,\,{\rm fm})^3$, so zero mode
effects in the masses are visible, and the simple linear fit clearly overshoots
the chiral limit point $m_f = -\mres$.}
\label{fig:pi-mf_b104}
\end{figure}

\begin{figure}
\begin{center}
\includegraphics[width=12cm]{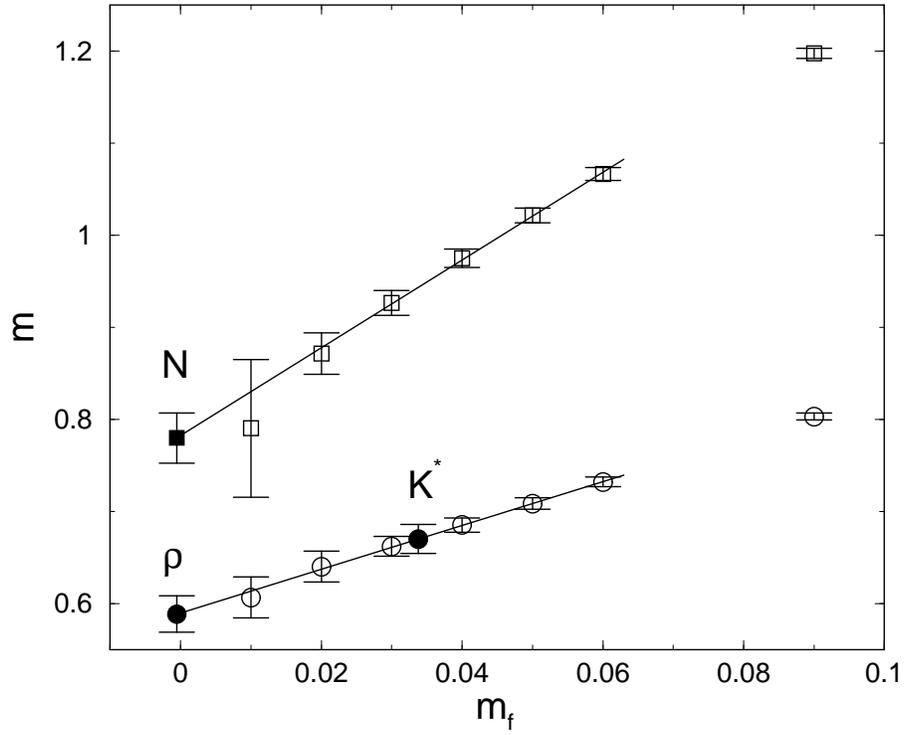}
\end{center}
\caption{Vector meson and baryon masses as functions of $m_f$ for
 $\beta=0.87$. Line is a linear fit to the data. The physical nucleon
 and $K^*$ mass are shown as solid points where $m_\rho$ is used to fix
 the scale.} 
\label{fig:rhoN-mf_b087}
\end{figure}

\begin{figure}
\begin{center}
\includegraphics[width=12cm]{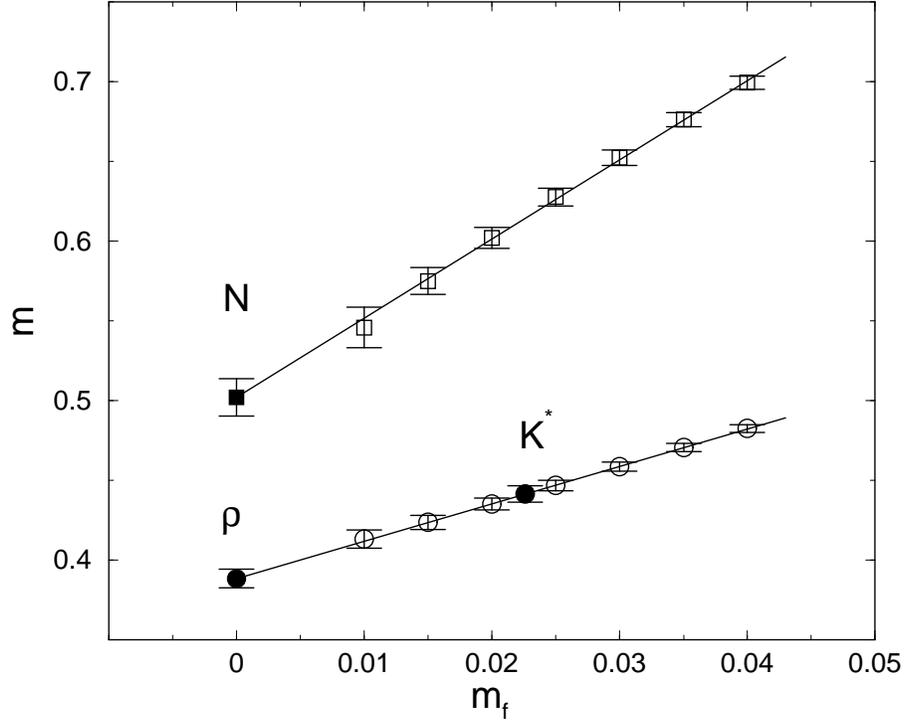}
\end{center}
\caption{Vector meson and baryon masses as functions of $m_f$ for
 $\beta=1.04$. The line is a linear fit to the data. The physical $K^*$
 mass is also shown.} 
\label{fig:rhoN-mf_b104}
\end{figure}

\begin{figure}
\begin{center}
\includegraphics[width=12cm]{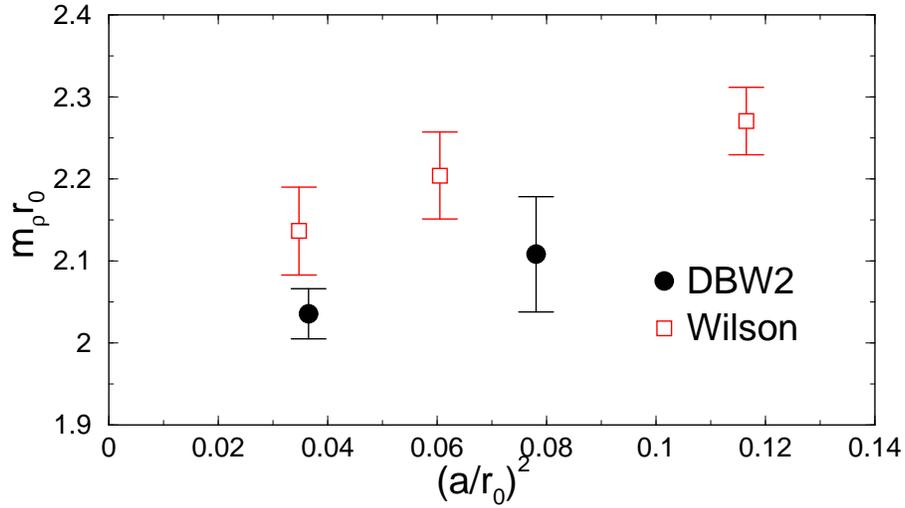}
\end{center}
 \caption{Scaling of the $\rho$ meson mass with the lattice spacing
 set by the Sommer parameter $r_0$. 
 The Wilson data are from ref.~\cite{Blum:2000kn}.}
\label{fig:rho-a}
\end{figure}

\begin{figure}
\begin{center}
\includegraphics[width=12cm]{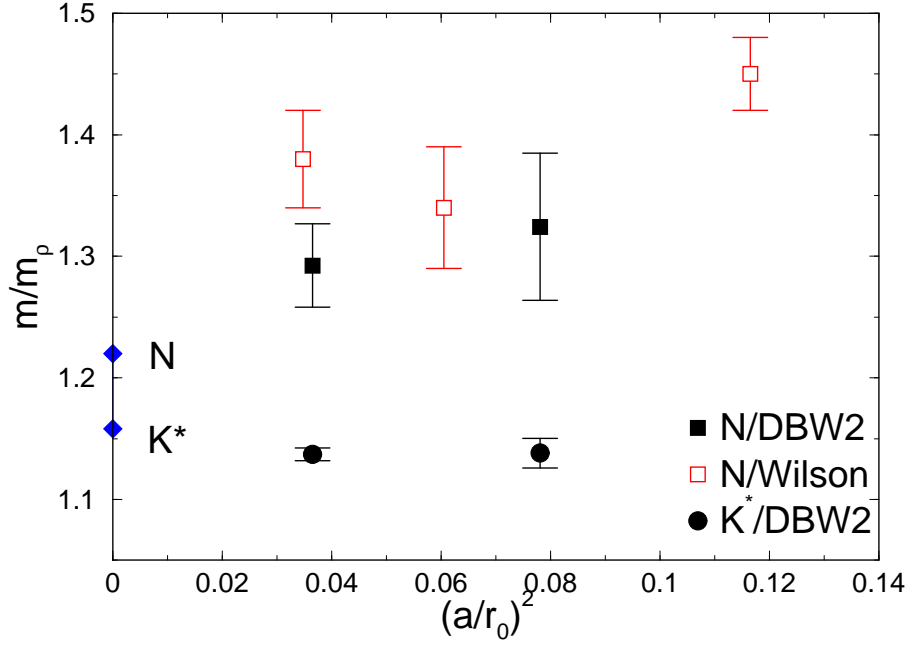}
\end{center}
 \caption{Scaling of the nucleon and $K^*$ masses with lattice
 spacing. The experimental mass of nucleon and $K^*$ are shown on the
 vertical axis. The Wilson data are from ref.~\cite{Blum:2000kn}.}
\label{fig:NKstar-a}
\end{figure}

\begin{figure}
\begin{center}
\includegraphics[width=12cm]{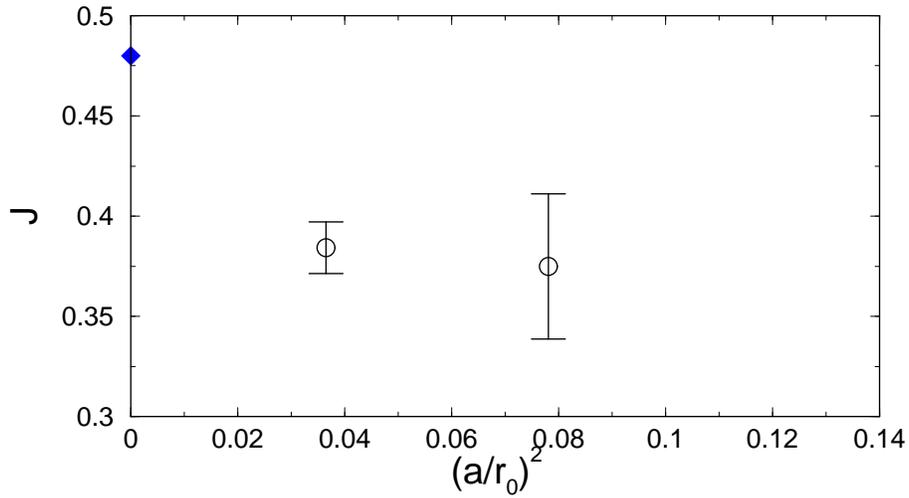}
\end{center}
 \caption{Scaling of the quenched J parameter with lattice spacing. The
 experimental point is shown on the vertical axis. The large
 discrepancy with experiment is due to quenching.}
\label{fig:J}
\end{figure}

\begin{figure}
\begin{center}
\includegraphics[width=12cm]{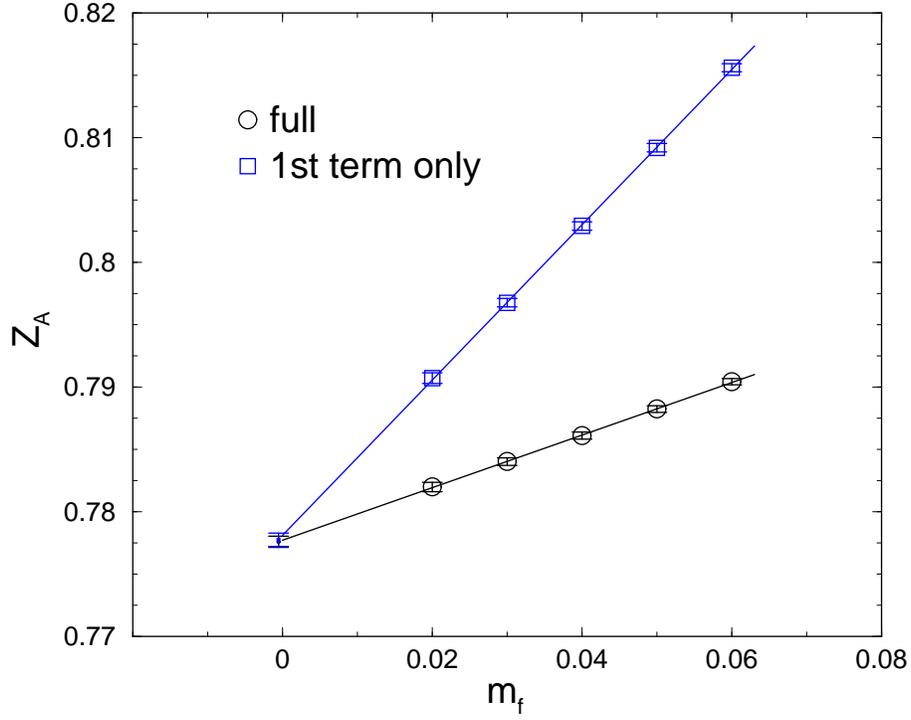}
\end{center}
 \caption{The local axial current renormalization factor as a function of
 $m_f$ for  $\beta=0.87$. The lines correspond to fits to Eq.~\ref{eq:R} 
 (lower)
 and the first term only (upper). The difference arises from lattice
 spacing errors that are proportional to $m_f$.}
\label{fig:Z_A-mf}
\end{figure}

\begin{figure}
\begin{center}
\includegraphics[width=12cm]{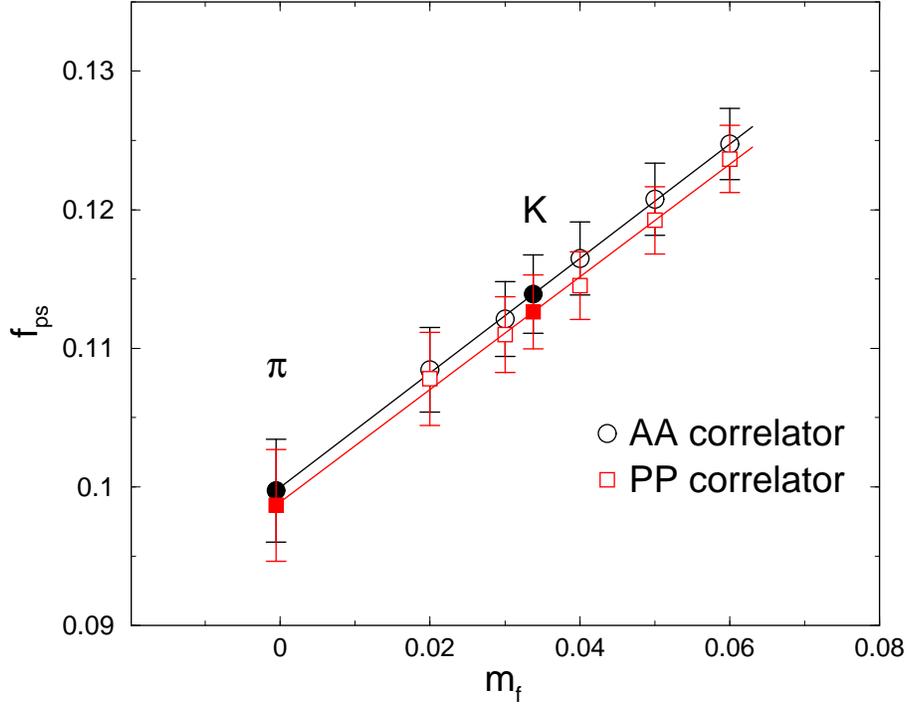}
\end{center}
 \caption{The quenched pseudoscalar decay constant as a function of $m_f$ for
 $\beta=0.87$. The results from the two techniques described in the
 text agree quite well. The data are fit to simple linear functions,
 and extrapolated (pion) or interpolated (kaon) to the physical points
 (filled symbols).
}
\label{fig:fps-mf_b087}
\end{figure}

\begin{figure}
\begin{center}
\includegraphics[width=12cm]{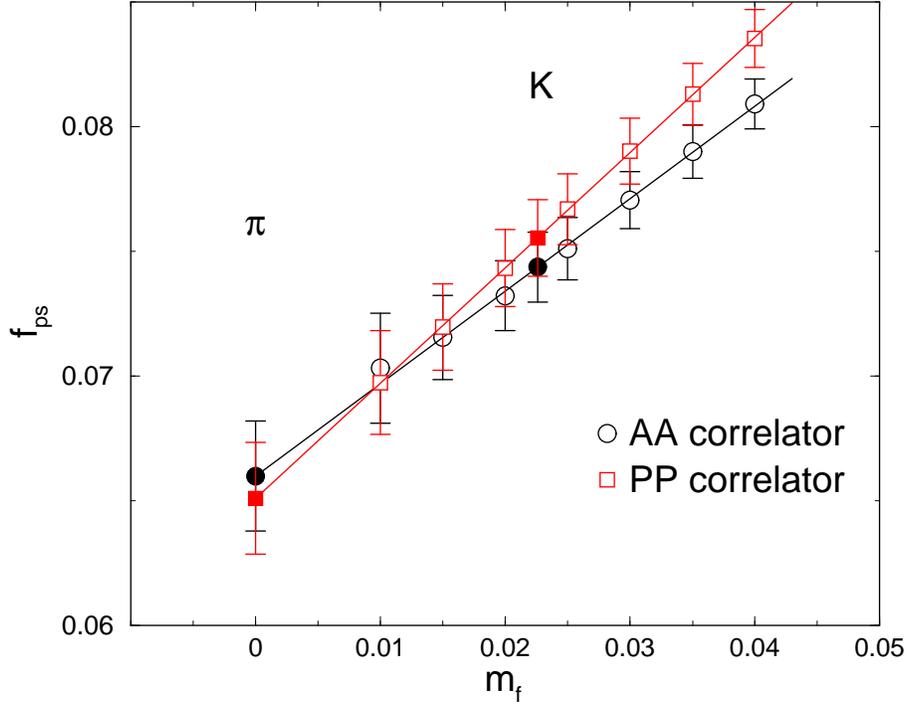}
\end{center}
 \caption{The same as Fig.~\ref{fig:fps-mf_b087}, but for
 $\beta=1.04$. Although the results agree within statistical errors,
 the agreement is not as good as that at $\beta=0.87$. Since the physical
 volume is significantly smaller in this case, we expect the difference is
 due to the presence of topological zero modes.}
\label{fig:fps-mf_b104}
\end{figure}

\begin{figure}
\begin{center}
\includegraphics[width=12cm]{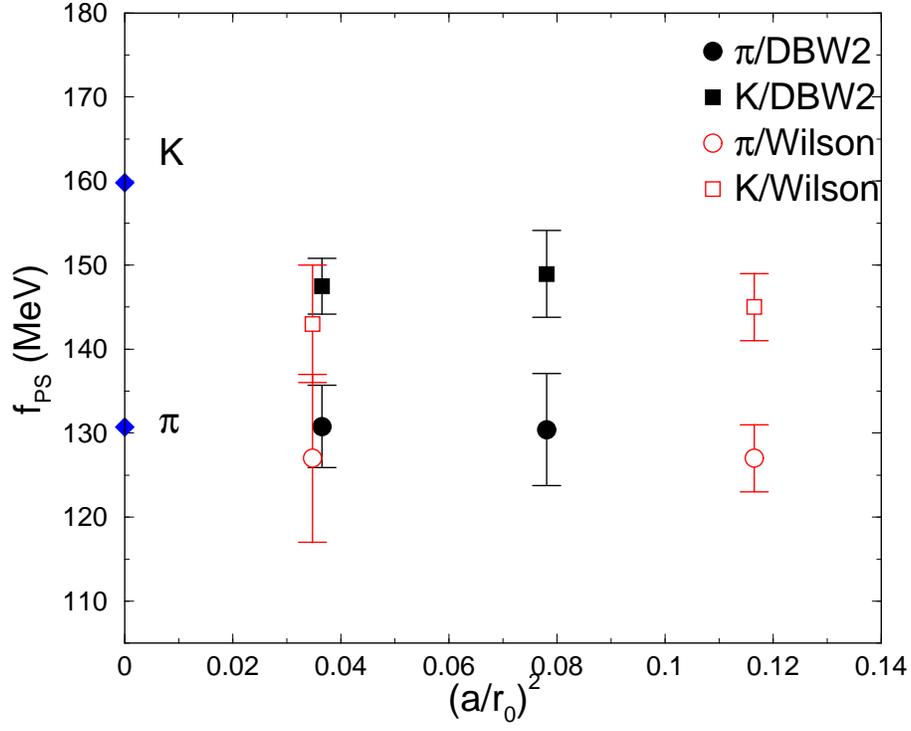}
\end{center}
 \caption{A scaling plot for the pion and kaon decay constants. The DBW2
 results appear to preserve the good scaling that was observed 
 with the use of the Wilson gauge action 
 (data from Ref.~\cite{Blum:2000kn})}.
\label{fig:fps-a}
\end{figure}

\fi

\end{document}